\documentclass[intlimits,twoside,a4paper]{article}
\usepackage{amssymb}
\usepackage{amsmath}
\usepackage{amsfonts}
\usepackage{graphicx}
\usepackage{multirow}
\usepackage{wrapfig}

\usepackage[T2A]{fontenc}
\usepackage[cp1251]{inputenc}

\usepackage[eqsecnum]{cmpj2}

\issuepage{165}

\issue{2012}{15}{2}{23802}

\doinumber{10.5488/CMP.15.23802}

\def\upA{\mathrm{A}}
\def\upB{\mathrm{B}}

\title[Modelling the ion-exchange equilibrium in nanoporous materials]%
{Modelling the ion-exchange equilibrium in nanoporous materials\thanks{Dedicated to Dr. Orest A. Pizio on the occasion of his
        60$^\mathrm{th}$ birthday.}}
\author[M. Luk\v{s}i\v{c}, V. Vlachy, B. Hribar-Lee]%
       {M. Luk\v{s}i\v{c}, V. Vlachy, B. Hribar-Lee}
\address{University of Ljubljana,
         Faculty of Chemistry and Chemical Technology,\\
         A\v{s}ker\v{c}eva c. 5,
         SI--1000 Ljubljana, Slovenia}

\authorcopyright{M. Luk\v{s}i\v{c}, V. Vlachy, B. Hribar-Lee, 2012}

\date{Received January 3, 2012, in final form February 2, 2012}
\begin{document}

\maketitle

\begin{abstract}
Distribution of a two component electrolyte mixture between the
model adsorbent and a bulk aqueous electrolyte solution was studied using
the replica Ornstein-Zernike theory and the grand canonical Monte
Carlo method. The electrolyte components were modelled to mimic the
HCl/NaCl and HCl/CaCl$_2$ mixtures, respectively. The matrix,
invaded by the primitive model electrolyte mixture, was formed from monovalent
negatively charged spherical obstacles. The solution was treated
as a continuous dielectric with the properties of pure water. Comparison of
the pair distribution functions (obtained by the two methods)
between the various
ionic species indicated a good agreement between the replica
Ornstein-Zernike results and machine calculations. Among
thermodynamic properties, the mean activity coefficient of the
invaded electrolyte components was calculated. Simple model for the
ion-exchange resin was proposed. The selectivity calculations yielded
qualitative agreement with the following experimental observations: (i)
selectivity increases with the increasing capacity of the adsorbent
(matrix concentration), (ii) the adsorbent is more selective for the ion
having higher charge density if its fraction in mixture is smaller.
\pacs 82.60.-s, 02.30.Rz, 61.20.-p
\end{abstract}

\section{Introduction}

A variety of phenomena of interest for basic science and technology
occur when an electrolyte solution or a mixture of electrolytes is in
equilibrium with a porous phase containing fixed charges.
Comprehension of the mechanism governing the equilibrium distribution of
ions between the two phases is therefore of considerable interest.
Materials containing charged micropores, such as membranes, glasses
and certain clays, have a property of partially rejecting an electrolyte
being filtered through them. In addition, such materials may be ion-selective;
the ions having different sizes or charges adsorb to a
different amount. This makes a basis for the separation process used
in water softening and purification, in treatment of radioactive wastes,
and for analytical purposes as well. Ion-exchange is, therefore, not only a
powerful tool in chemical analysis and in scientific research, but also of
great importance for everyday life \cite{Helferich,Zagorodni,Boda1,Boda2}.

For a quantitative description of these phenomena, some knowledge of
the spatial distribution of the counter-ion and co-ion species in the
micropores is needed. To begin a mathematical description, a
suitable model of the adsorbing material should first be assumed. The
ion-exchange resins are quite often modelled as spherical holes or
cylindrical channels. In such cases the calculations are, due to the
known geometry, relatively simple (see, for example, \cite{Vlachy1990})
and the Poisson-Boltzmann theory
is most often accurate
enough to make viable predictions of measurable properties.
Alternatively, the Monte Carlo or Brownian Dynamics methods may be
used for this
purpose \cite{Jamnik1993,Jamnik1995,Hribar2000a,Jardat2008,Jardat2012}.

In contrast to the models with well defined (spherical, cylindrical,
etc.) geometry described above, we are here interested in modelling the porous
systems where distribution of charges in a system is not governed
by a simple geometry. The porous material (matrix) is pictured as a
set of charged obstacles. Distribution of the
obstacles is obtained by a certain recipe (see, for example,  \cite{Luksic2007,Jardat2008}). Inhomogeneity that causes the
field in which mobile ions are distributed is given on a molecular
level. Such a model is a representative of a quenched-annealed
systems which have been more extensively studied only in the last 25
years. The progress in this area of research is documented in several
review papers
\cite{pizio1998,Rosinberg1999,Pizio2000,Hribar2010a}.

The pioneering contributions to the science of partly quenched
systems are due to Madden, Glandt and coworkers
\cite{madden1992, madden1988,fanti1990,ford1994}, Given and Stell
\cite{given1995, given1992, given1994, given1993}, Rosinberg
\cite{rosinberg1994}, Bratko and Chakraborty
\cite{Bratko1995,Bratko1996}, Chandler and coworkers
\cite{Deem1994,leung1995}, Kaminsky and Monson
\cite{kaminsky1991,vega1993}, Holovko and coworkers
\cite{holovko1,holovko2}, to mention only the most visible contributors.
In references \cite{madden1992, madden1988} and \cite{given1995,
given1992, given1994, given1993} the replica Ornstein-Zernike
theory was proposed to calculate the thermodynamic properties and
spatial distribution of particles in partly quenched systems with
short-range forces.

Following these pioneering studies, Pizio and coworkers
\cite{Hribar1997} succeeded in extending the replica theory to
the systems with Coulomb interactions. In a series of papers
\cite{hribar1998,hribar1999,hribar2000,hribar2001,hribar2002,Vlachy2004}
the quenched system was modelled as a set of charged obstacles
containing positive and negative charges so that the matrix subsystem (the
set of obstacles) was electroneutral. The other electroneutral
subsystem, that is electrolyte solution, was assumed to occupy the
void space in-between the obstacles, i.e., to invade the matrix. To
calculate the thermodynamic properties and spatial distribution
between the quenched and annealed particles the replica
Ornstein-Zernike (ROZ) integral equations were applied. In addition, an
expression was derived \cite{hribar2002} for the mean activity coefficient valid within the replica
hypernetted-chain approximation.
The canonical and grand canonical Monte Carlo method was
used to verify the theoretical approximations inherent to the replica
theories. Very good agreement between the theory and simulations
was obtained for both, thermodynamic parameters and for the pair
distribution functions (see, for example, \cite{hribar2001,hribar2002,Vlachy2004}). More recently, good
agreement was confirmed by the independent Brownian Dynamics
simulations based on the same model \cite{Jardat2008}.

In the last years, the model for a quenched system was modified to allow
studying the ion-exchange phenomena. The model proposed in
references \cite{Luksic2007,Luksic2007a,Luksic2009a} differs from
the one presented above in one important aspect. We assume again
that the distribution of charged obstacles corresponds to an
equilibrium distribution of ions in an electrolyte solution under
conditions given by the temperature and the dielectric constant before
quenching.  However, in contrast to the previous studies, we assume
that only anions are quenched and the cations are allowed to
equilibrate (anneal). We also assume that the model system is in
thermodynamic equilibrium with an external electrolyte solution, which
penetrates into the matrix. A similar model, but for a somewhat simpler
system, was previously studied by Pastore et al. \cite{Pastore1996}.
The model was recently studied by the grand canonical Monte Carlo
method \cite{Luksic2007,Luksic2007a} as well as the Brownian
Dynamics approach \cite{Jardat2012}. The simulations indicate that the
ROZ theory in the hypernetted-chain approximation, adapted to this
model in \cite{Luksic2007}, yields very good agreement with
computer simulations.

In the present article, we extend the approach of the
previous studies \cite{Luksic2007,Luksic2007a} to the investigation of  the
ion-selectivity effect in the matrix representing a charged nanoporous
system. The system of quenched obstacles  formed from the primitive
model ions is negatively charged. Within this system (matrix) the
annealed ions are distributed; an excess of cations is present to
maintain the electroneutrality condition. In equilibrium, the
concentration of the invading electrolyte is determined by the
properties of the bulk solution of the same chemical composition. Such a
system can represent a crude model of the ion-exchange resin. The
distribution of ionic species within the matrix is calculated by the ROZ
integral equations and in several cases also by the grand canonical
Monte Carlo (GCMC) method. For a certain composition of the external
electrolyte, a different equilibrium composition of the internal
electrolyte solution is established, favouring the ionic species, which
interact stronger with the matrix particles. Structural, as well as the
thermodynamic properties are calculated, such as the mean activity or Donnan
exclusion coefficients, varying the composition of an
external electrolyte. The ion-exchange isotherms, reflecting the ion
selectivity in HCl/NaCl and HCl/CaCl$_2$ mixtures, are presented
as functions of the model parameters.

\clearpage

\section{The model and theoretical methods}

To model an ion-exchange phenomenon, a quenched-annealed system composed of two subsystems
was considered. The quenched
negatively charged ionic obstacles represented the matrix subsystem,
while the annealed subsystem was a mixture of two primitive model
electrolytes with the addition of cations needed to compensate the
negative charge of the matrix. It was assumed that the matrix
remained insensitive to the presence of the annealed fluid, i.e., its
structure did not change upon the introduction of the annealed phase.
The solvent was treated as a structureless  continuum (McMillan-Mayer
level of description) characterized by the dielectric constant of pure
water under the conditions of the study. Throughout the paper,
indices 0 and 1 designate the matrix and the annealed subsystems,
respectively.

The matrix phase was prepared in the following way: the $+1:-1$
restricted primitive model electrolyte, with the diameters of ions
$\sigma_+^0 = \sigma_-^0 = 4.25$~\AA, was quenched at a certain
temperature $T_0$, where the dielectric constant of the medium was
$\varepsilon_0$. The structure of the matrix was considered to
represent one of the equilibrium structures, corresponding to a given
set of conditions (concentration, temperature, relative permittivity), and
governed by the Coulomb pair interaction potential
\begin{equation}
\beta_0 U_{ij}^{00} = \left \{
\begin{array}{cc}
\infty, & r < (\sigma_i^0 + \sigma_j^0)/2,\\
z_i^0 z_j^0 \lambda_{\mathrm{B},0}/r, & r \geqslant (\sigma_i^0 + \sigma_j^0)/2.
\end{array}
\right .
\label{eq:pot-matrix}
\end{equation}
In equation \eqref{eq:pot-matrix}, $z_i^0$ ($z_j^0$) is the charge of the
ion $i$ ($j$), equal to $+1$ or $-1$, $\lambda_{\mathrm{B},0} =
e_0^2 / (4\pi\varepsilon_v \varepsilon_0 k_\mathrm{B} T_0)$ is the
Bjerrum length ($e_0$ denotes the unit charge, $\varepsilon_v$ the
permittivity of vacuum, and $k_\mathrm{B}$ the Boltzmann constant),
$r$ designates the distance between the centres of ions $i$ and $j$,
and $\beta_0 = 1/(k_\mathrm{B}T_0)$. In the final stage of the matrix
preparation, the cations were disregarded, so that the quenched
phase consisted only of a spatially fixed arrangement of negatively
charged ($z_{-}^0 = -1$) particles. Conditions of the matrix preparation
corresponded to $\lambda_{\mathrm{B},0} = 7.14$~\AA\ (water
solution at 298 K), and ion concentrations $c_0 = 0.1$, $0.5$, $1.0$, and
$2.0$~mol\,dm$^{-3}$.

The annealed subsystem was modelled as a mixture of two $+1:-1$
electrolytes, or a mixture of a $+1:-1$ and $+2:-1$ electrolyte having a
common anion.  An additional number of univalent cations (common
to cations of one of the electrolytes in the mixture), was present to
neutralize the overall negative charge of the matrix. The diameters of
the $+1$ ions were $\sigma_+^1 = 5.04$~\AA\ (model for hydrogen
ions, H$^{+}$) or $\sigma_+^1 = 3.87$~\AA\ (model for sodium ions,
Na$^{+}$), of the $+2$ ions $\sigma_+^1 = 7.03$~\AA\ (model for
calcium ions, Ca$^{2+}$), and of the $-1$ ions $\sigma_{-}^1 =
3.63$~\AA\ (model for chloride ions, Cl$^{-}$). Models of HCl and
NaCl, or HCl and CaCl$_2$ mixtures were, therefore, considered.
Cations compensating the matrix charge were taken to be H$^{+}$,
mimicking in this way a H$^{+}$-ion-exchange resins. The system
was considered to thermally equilibrate under the conditions of $T_1$
and $\varepsilon_1$, with the pair potential for a $0-1$ interaction
being equal to
\begin{equation}
\beta_1 U_{-i}^{01} = \left \{
\begin{array}{cc}
\infty, & r < (\sigma_{-}^0 + \sigma_i^1)/2,\\
z_{-}^0 z_i^1 \lambda_{\mathrm{B},1}/r, & r \geqslant (\sigma_{-}^0 + \sigma_i^1)/2,
\end{array}
\right .
\label{eq:pot-matrixannealed}
\end{equation}
and for a $1-1$ interaction
\begin{equation}
\beta_1 U_{ij}^{11} = \left \{
\begin{array}{cc}
\infty, & r < (\sigma_i^1 + \sigma_j^1)/2,\\
z_i^1 z_j^1 \lambda_{\mathrm{B},1}/r, & r \geqslant (\sigma_i^1 + \sigma_j^1)/2.
\end{array}
\right .
\label{eq:pot-annealed}
\end{equation}
In equations \eqref{eq:pot-matrixannealed} and
\eqref{eq:pot-annealed}, $\lambda_{\mathrm{B},1} = e_0^2 /
(4\pi\varepsilon_v \varepsilon_1 k_\mathrm{B} T_1)$ is the Bjerrum
length for the annealed solution, and $\beta_1 =
1/(k_\mathrm{B}T_1)$. Although the conditions for the preparation of
the quenched phase can differ from the conditions of observation, we
assumed them to be equal in this study, i.e. $\lambda_{\mathrm{B},1}
= \lambda_{\mathrm{B},0} = 7.14$~\AA.

\subsection{The replica Ornstein-Zernike integral equation theory}

A case of adsorption of a single electrolyte within the charged matrix
was considered by our group several years ago \cite{Luksic2007}.
Here, we extend the theoretical description to the study of mixtures of
electrolytes with a common ion.

In order to define the spatial distribution of matrix ions we first need to
obtain the structure for a $+1:-1$ electrolyte. This is  obtained by
solving a set of Ornstein-Zernike (OZ) equations in the form
\begin{equation}
\begin{bmatrix}
h_{++}^{00} & h_{+-}^{00} \\
h_{-+}^{00} & h_{--}^{00}
\end{bmatrix}
=
\begin{bmatrix}
c_{++}^{00} & c_{+-}^{00} \\
c_{-+}^{00} & c_{--}^{00}
\end{bmatrix}
+
\begin{bmatrix}
c_{++}^{00} & c_{+-}^{00} \\
c_{-+}^{00} & c_{--}^{00}
\end{bmatrix}
\otimes
\begin{bmatrix}
\rho_{+}^{0} & 0 \\
0 & \rho_{-}^{0}
\end{bmatrix}
\cdot
\begin{bmatrix}
h_{++}^{00} & h_{+-}^{00} \\
h_{-+}^{00} & h_{--}^{00}
\end{bmatrix}
\, ,
\label{eq:matrix}
\end{equation}
where $h$ and $c$ stand for the total and the direct correlation
functions, respectively, and $\otimes$ denotes the convolution in
$r$-space. Due to spherical symmetry, $h_{+-}^{00} = h_{-+}^{00}$
and $c_{+-}^{00} = c_{-+}^{00}$. Equations \eqref{eq:matrix} need to
be renormalized before being solved numerically (see, for example, \cite{duh1992,ichiye1990}). The hypernetted-chain (HNC)
closure relation between $h$ and $c$ ($c_{ij} = h_{ij} - \ln \left [ h_{ij}
+ 1 \right ] - \beta_0 U_{ij}^{00}$) was used to obtain the solution. Note that
only $h_{--}^{00}$ is needed in the subsequent replica OZ equations
(\ref{eq:roz1}). This function contains all the information of
the matrix subsystem.

The replica Ornstein-Zernike equations for the quenched-annealed system
(cf. \cite{given1994,Luksic2009a}) read
\begin{align}
h_{i-}^{10} &= c_{i-}^{10} + c_{i-}^{10} \otimes \rho^0_- h_{--}^{00} +
  \sum_{k = \upA,\upB,-} \left [ c_{ik}^{11} \otimes \rho_k^1 h_{k-}^{10} -
   c_{ik}^{12} \otimes \rho_k^1 h_{k-}^{10} \right ] , \label{eq:roz1}\\
h_{ij}^{11} &=  c_{ij}^{11} + c_{i-}^{10} \otimes \rho^0_{-} h_{-j}^{01} +
   \sum_{k = \upA,\upB,-} \left [ c_{ik}^{11} \otimes \rho_k^1 h_{kj}^{11} -
    c_{ik}^{12} \otimes \rho_k^1 h_{kj}^{21} \right ] , \label{eq:roz2}\\
h_{ij}^{12} &= c_{ij}^{12} + c_{i-}^{10} \otimes \rho^0_{-} h_{-j}^{01} +
   \sum_{k = \upA,\upB,-} \left [   c_{ik}^{11} \otimes \rho_k^1 h_{kj}^{12} +
      c_{ik}^{12} \otimes \rho_k^1 h_{kj}^{11} -
     2c_{ik}^{12} \otimes \rho_k^1 h_{kj}^{21} \right ] ,
\label{eq:roz3}
\end{align}
where indices $i$ and $j$ stand for cations or anions of the annealed
electrolytes in the mixture (i.e., H$^+$, Na$^+$  and Cl$^-$, or H$^+$,
Ca$^{2+}$, and Cl$^-$), while A and B denote only the cations of the
annealed mixture (H$^+$ and Na$^+$, or H$^+$ and Ca$^{2+}$,
respectively). We stress that the $\rho_\mathrm{H^+}^1$ is the total
density of H$^+$ ions in the system which belong to the HCl and to
the cations compensating the negative charge of the matrix.

As in the case of matrix preparation (see expression (\ref{eq:matrix})),
equations \eqref{eq:roz1}--\eqref{eq:roz3} need to be renormalized
prior to numerical solution (i.e. $c$ and $h$ need to be split into a short and long range part). The procedure is documented in \cite{hribar1998,hribar1999}. The only difference from the published
equations is that here $\kappa^2$ contains the sum over all the
annealed species. Additionally, HNC closure condition was applied to
correlate the $h$ and $c$ functions:
\begin{align}
c_{ij}^{mn} &= h_{ij}^{mn} - \ln \left [ h_{ij}^{mn} + 1 \right ] - \beta_1 U_{ij}^{mn}  ,  \label{eq:hncr1} \\
c_{ij}^{12} &= h_{ij}^{12} - \ln \left [ h_{ij}^{12} + 1 \right ]  ,
\label{eq:hncr2}
\end{align}
where $m$ and $n$ take the values $0$ and $1$, and $i$ ($j$) in
equation (\ref{eq:hncr1}) corresponds to the annealed cation or anions
(H$^+$, Na$^+$, Ca$^{2+}$, and Cl$^-$) and to the matrix ions. $i$
($j$) indexes in equation (\ref{eq:hncr2}) belong only to annealed
ions. Note that $U^{12}_{ij} = 0$ since there is no interaction
between the ions of different replicas. Equations
\eqref{eq:roz1}--\eqref{eq:roz3} with the HNC closure were solved by a
direct iteration on a grid of $2^{15}$ points with $\Delta r = 0.005$~\AA.

The individual activity coefficient within the replica HNC formalism
reads \cite{hribar2000}
\begin{eqnarray}
\ln \gamma _i ^1 &=&- \,\rho _{-} ^0 \int c_{(\mathrm{s})i-}^{10} \mathrm{d}{\bf r}
                    -\sum_{j=\upA,\upB,-} \rho_j^1 \int \left [ c_{(\mathrm{s})ij}^{11} -
                     c_{(\mathrm{s})ij}^{12} \right ] \mathrm{d}{\bf r}  \nonumber \\
&&{}+ \frac{\rho_{-}^0}{2} \int h_{i-}^{10} \left ( h_{i-}^{10} - c_{i-}^{10} \right )
\mathrm{d}{\bf r}
+ \frac{1}{2} \sum_{j=\upA,\upB,-} \rho_j^1 \int
\left [ h_{ij}^{11} \left ( h_{ij}^{11} - c_{ij}^{11} \right ) - h_{ij}^{12} \left (
h_{ij}^{12} - c_{ij}^{12} \right ) \right ] \mathrm{d}{\bf r},
\label{eq:gammai}
\end{eqnarray}
where $c_\mathrm{(s)}$ denotes the short range part of the direct
correlation function coming from the renormalization procedure, and $\mathrm{d}{\bf r} = 4\pi r^2 \mathrm{d}r$.
Again, A and B denote cations in the mixture (H$^+$ and Na$^+$, or
H$^+$ and~Ca$^{2+}$).


\subsection{Monte Carlo simulations}

The matrix was prepared from a size symmetric ($\sigma_+^0 =
\sigma_-^0 = 4.25$~\AA) $+1:-1$ electrolyte using a separate
canonical Monte Carlo simulation. After equilibration, the matrix
anions were frozen in their positions, while the cations were replaced
by the model hydrogen cations ($\sigma_\mathrm{H^+} = 4.25$~\AA)
and made it possible to move freely, equilibrating together with the annealed
electrolyte mixture (HCl/NaCl or HCl/CaCl$_2$). The annealed electrolyte ions were then distributed
within the matrix and the system was studied using the Monte Carlo method in the
grand canonical ensemble. The methodology of this approach is well
established and extensively described in several previous papers and
is not, therefore, repeated here
\cite{Luksic2007,Luksic2007a,Luksic2009a}.

The details of simulations are as follows: The number of matrix
particles in the simulation box varied from 300 to 2500, depending on the
concentration. The average number of fluid cation species
distributed within the matrix varied from 500 to 3500. The periodic boundary conditions within the minimum image convention were applied. The ions within
the matrix were first equilibrated over at least $10^{7}$ Monte Carlo steps. After the equilibration, a production
run of $2\cdot 10^{8}$ attempted configurations was carried out to
obtain the average concentration of the adsorbed electrolyte species. Note
that thermodynamic properties were averaged over different annealed
fluids, as well as over one to two different matrix configurations.

To calculate the equilibrium concentration in the bulk electrolyte, the
activity coefficients of the bulk electrolyte mixture should be known in
advance. These values were obtained using the hypernetted-chain
(HNC) theory which has proved to be very successful in describing
the properties of ionic fluids~\cite{gutierrez}.

\section{Results and discussion}

The two theoretical methods described above were established to be
complementary in the sense that they yield consistent results for
structural and thermodynamic properties of electrolyte solutions
adsorbed in nanoporous materials
\cite{hribar2001,hribar2002,Luksic2007,Luksic2007a,Luksic2009a}.
While the ROZ/HNC theory is computationally less demanding, and is,
therefore, faster to use for systematic investigations of different
effects on the system properties, the GCMC method is more
convenient in studying the equilibrium distribution of ions between
the bulk and matrix phase. In this paper we used both methods; the
results are organized as follows.
First, we tested the performance of
the ROZ/HNC theory for the case studied here by comparing the
structural (pair distribution functions) and thermodynamic properties
(mean activity coefficients, $\gamma_\pm^{\nu_+ + \nu_-} = \gamma_+^{\nu_{+}}\gamma_-^{\nu_{-}}$) obtained using the two methods. Next, the
effect of the matrix concentration (e.g. adsorbent capacity) on
\begin{table}[ht]
\caption{The mean activity coefficients of electrolytes
adsorbed in the matrix:  $\gamma_\pm$, obtained from the
ROZ/HNC theory and GCMC simulation for different matrix
concentrations. Concentrations of HCl and NaCl
($c_\mathrm{HCl}$ and $c_\mathrm{NaCl}$, respectively) and of the
matrix ($c_0$) are given in mol\,dm$^{-3}$. $I^\mathrm{out}=0.5$~mol\,dm$^{-3}$, $\lambda_{\mathrm{B}, 0}
= \lambda_{\mathrm{B}, 1} = 7.14$~\AA. The numerical error of both methods was
estimated to be in the last digit given.} \label{tbl:hclnacl}
\vspace{2ex}
\footnotesize{
\begin{center}
\begin{tabular}{cccccccccc}
\hline\noalign{\smallskip}
\multicolumn{5}{c}{} & \multicolumn{2}{c}{$\gamma_\pm^\mathrm{HCl}$} & &
\multicolumn{2}{c}{$\gamma_\pm^\mathrm{NaCl}$}\\\noalign{\smallskip}
\hline\noalign{\smallskip}
$c_0$ & & $c_\mathrm{HCl}$ & $c_\mathrm{NaCl}$ & & GCMC & ROZ & & GCMC & ROZ\\\noalign{\smallskip}
\hline\noalign{\smallskip}
\multirow{7}{*}{1.0}
& & 0.107	& 0.0414 & & 1.039  & 1.072 & & 0.901 & 0.928\\
& & 0.0845	& 0.0825 & & 1.042  & 1.074 & & 0.900 & 0.928\\
& & 0.0646	& 0.123  & & 1.046  & 1.077 & & 0.901 & 0.929\\
& & 0.0473	& 0.163  & & 1.051  & 1.081 & & 0.903 & 0.930\\
& & 0.0325	& 0.202  & & 1.057  & 1.086 & & 0.906 & 0.932\\
& & 0.0206	& 0.240  & & 1.064  & 1.092 & & 0.909 & 0.934\\
& & 0.0114	& 0.277  & & 1.072  & 1.100 & & 0.913 & 0.938\\\noalign{\smallskip}
\hline\noalign{\smallskip}
\multirow{7}{*}{0.5}
                 & &     0.227	 & 0.0464    & &  0.879    &  0.888   & & 0.804  &   0.807 \\
                 & &     0.188	 & 0.0930    & &  0.876    &  0.885   & & 0.798  &   0.804 \\
                 & &     0.151	 & 0.139     & &  0.874    &  0.883   & & 0.795  &   0.802 \\
                 & &     0.117	 & 0.186     & &  0.873    &  0.881   & & 0.792  &   0.800 \\
                 & &     0.0845  & 0.231     & &  0.872    &  0.880   & & 0.791  &   0.798 \\
                 & &     0.0562  & 0.277     & &  0.872    &  0.880   & & 0.789  &   0.797 \\
                 & &     0.0325  & 0.321     & &  0.874    &  0.881   & & 0.789  &   0.796 \\\noalign{\smallskip}
\hline\noalign{\smallskip}
\multirow{9}{*}{0.1}
                & &      0.396  & 0.0494	    & & 0.805    & 0.806   & & 0.755   &  0.754  \\
                & &      0.347  & 0.0990	    & & 0.801    & 0.802   & & 0.750   &  0.750  \\
                & &      0.298  & 0.148  	    & & 0.796    & 0.797   & & 0.746   &  0.747  \\
                & &      0.249  & 0.198	            & & 0.793    & 0.793   & & 0.743   &  0.743  \\
                & &      0.201  & 0.247             & & 0.788    & 0.789   & & 0.739   &  0.739  \\
                & &      0.153  & 0.297 	    & & 0.784    & 0.784   & & 0.735   &  0.736  \\
                & &      0.106  & 0.346 	    & & 0.780    & 0.781   & & 0.732   &  0.732  \\
                & &      0.0602 & 0.396 	    & & 0.777    & 0.777   & & 0.729   &  0.729  \\
                & &      0.0200 & 0.445             & & 0.774    & 0.774   & & 0.726   &  0.726  \\
\noalign{\smallskip}
\hline\noalign{\smallskip}
\end{tabular}
\end{center}}
\end{table}
the
mean activity coefficient, which determines the chemical equilibrium
in the system, was examined using the ROZ/HNC theory. The Donnan
exclusion coefficients and the ion-exchange isotherms, reflecting the
selectivity of an adsorbent were evaluated by the Monte Carlo simulation
in the grand canonical ensemble.

\subsection{Comparison of the ROZ/HNC results with computer simulations}

\begin{figure}[ht]
\centerline{\includegraphics[width=5.5cm]{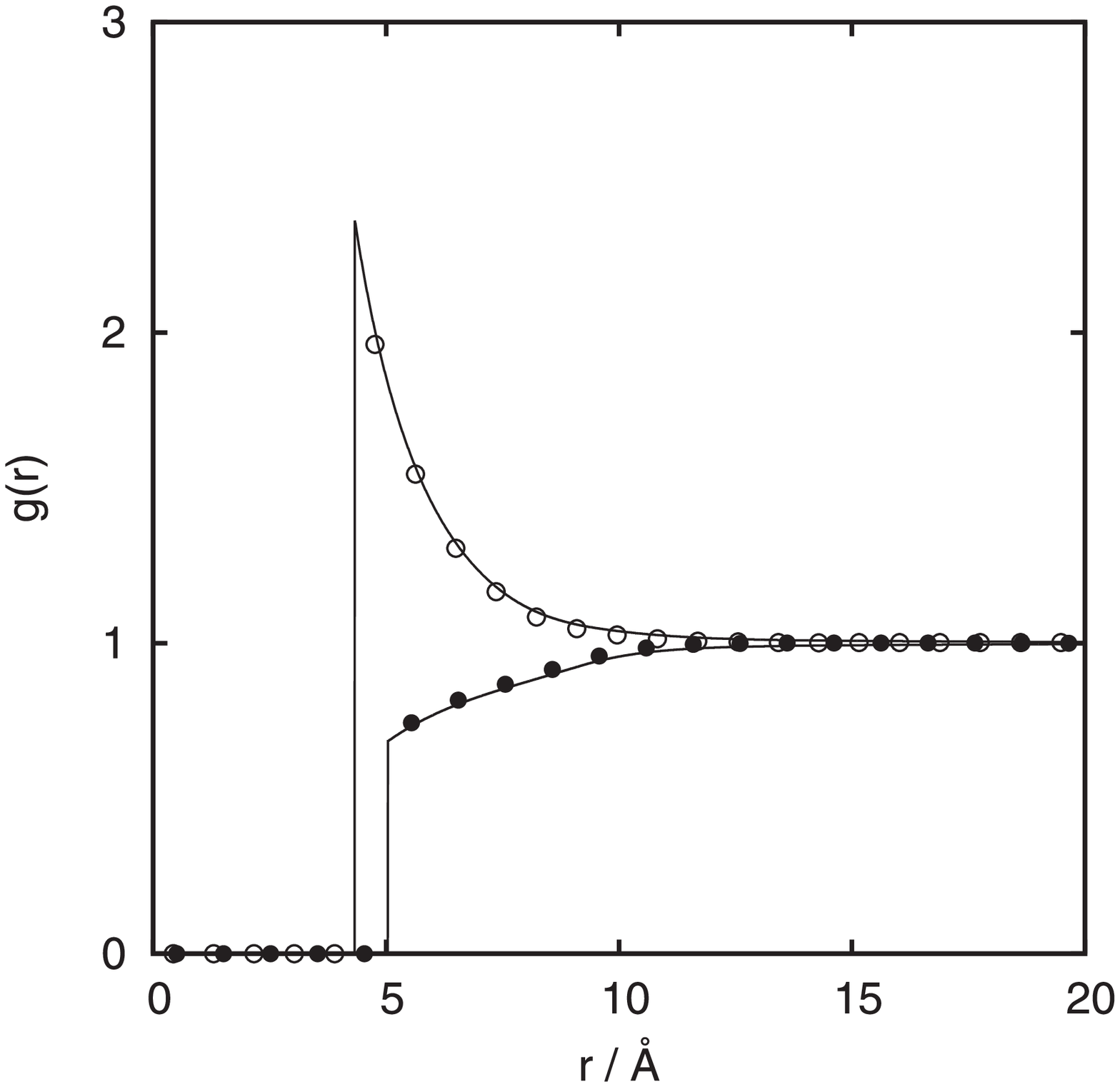}\hspace{1cm}
\includegraphics[width=5.5cm]{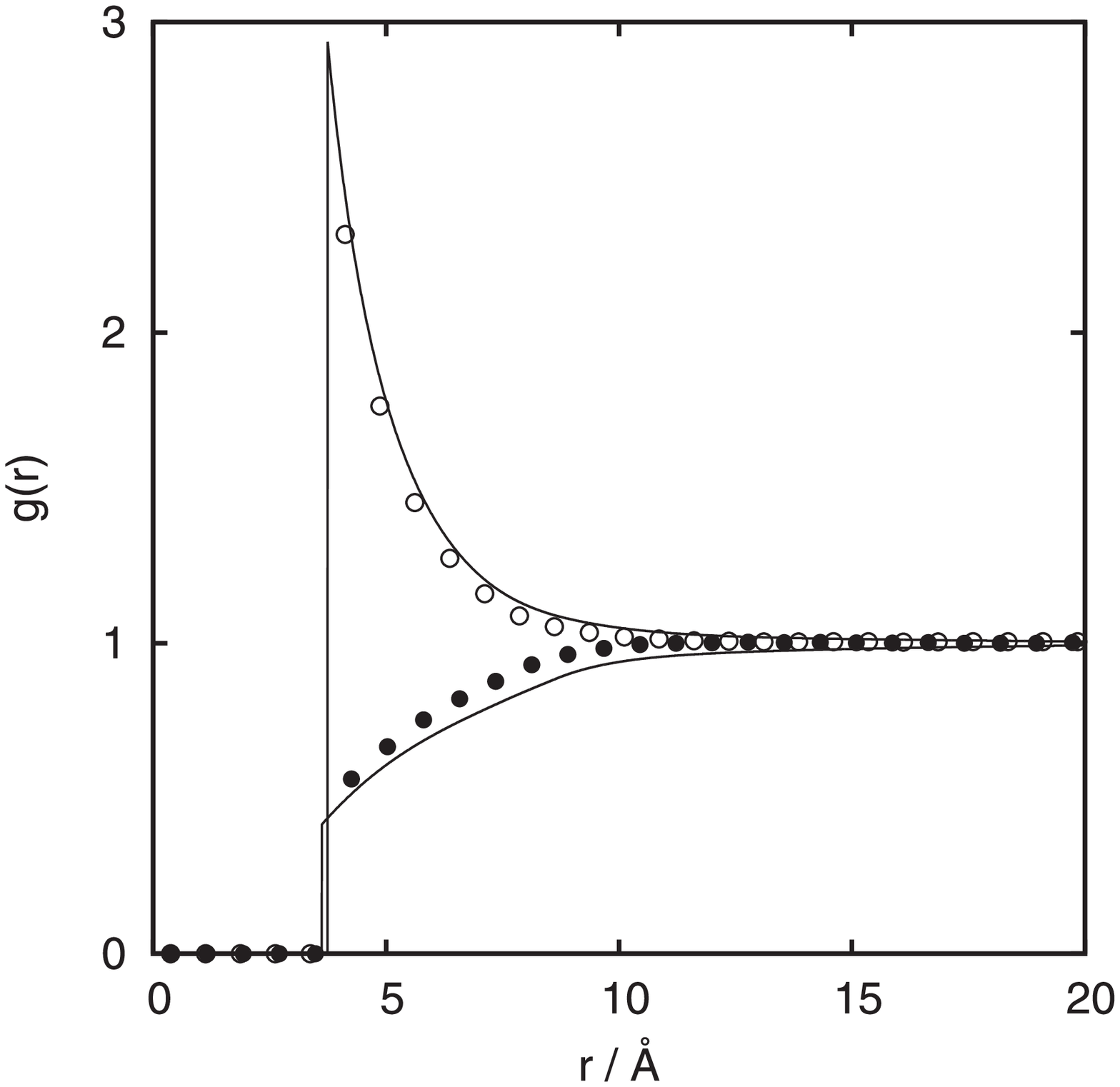}}
\caption{Fluid-fluid pair distribution functions (mixtures of HCl and NaCl).
Left: H$^+$--H$^+$ ($\medbullet$) and
H$^+$--Cl$^-$ ($\medcirc$); Right: Na$^+$--Na$^+$  ($\medbullet$) and
Na$^+$--Cl$^-$  ($\medcirc$). $c_0 =
1.0$~mol\,dm$^{-3}$, $c_\mathrm{HCl}^\mathrm{in} = 0.0325$~mol\,dm$^{-3}$,
$c_\mathrm{NaCl}^\mathrm{in} =
0.2018$~mol\,dm$^{-3}$. Symbols represent GCMC data while lines show the
ROZ/HNC theory. $\lambda_\mathrm{B,0} = \lambda_\mathrm{B,1} = 7.14$~\AA.}
\label{fig:gr-na}
\end{figure}

\begin{figure}[!h]
\centerline{\includegraphics[width=5.5cm]{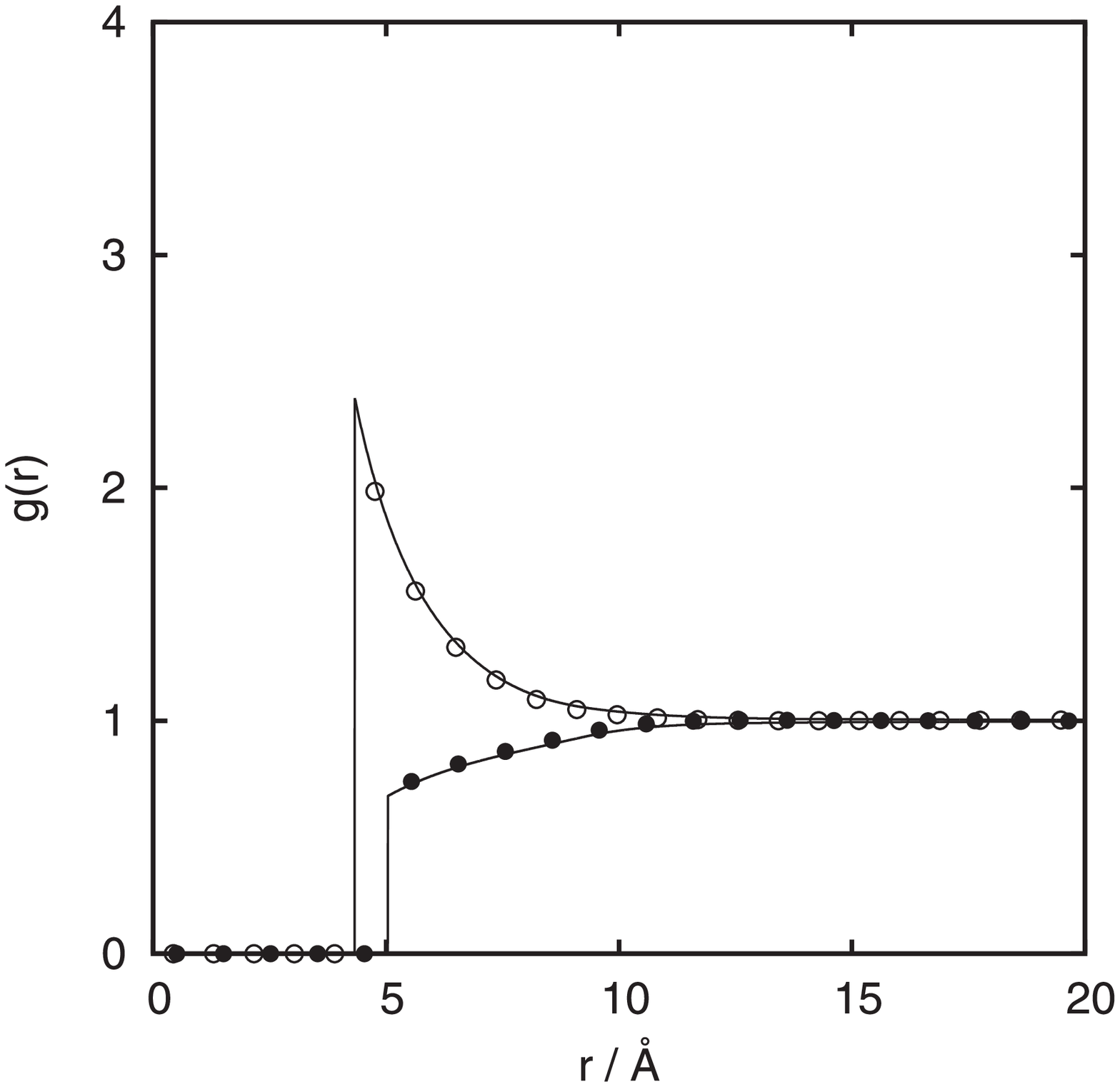}\hspace{1cm}
\includegraphics[width=5.5cm]{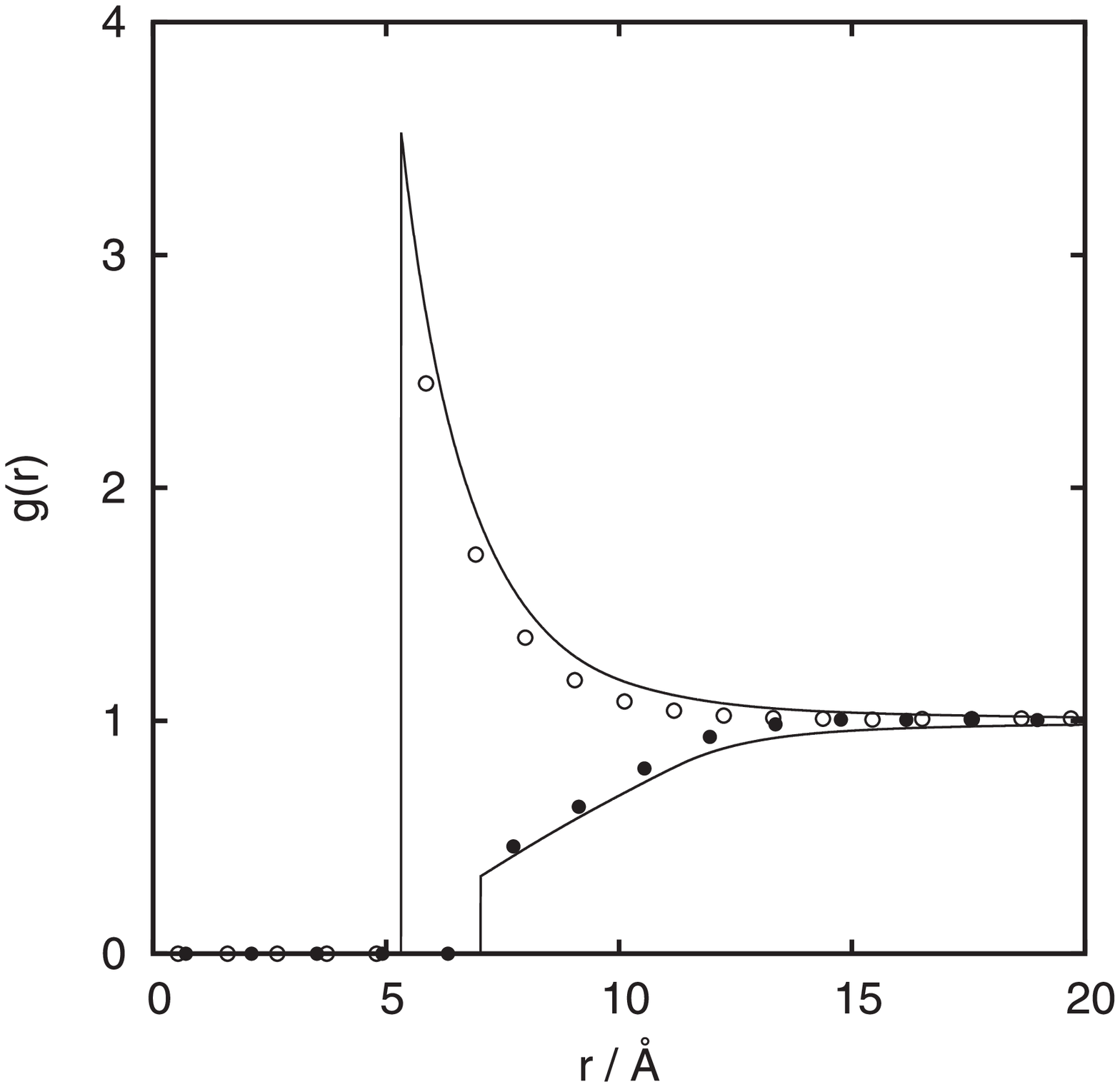}}
\caption{Fluid-fluid pair distribution functions (mixtures of HCl and CaCl$_2$). Left: H$^+$--H$^+$ ($\medbullet$)
and H$^+$--Cl$^-$  ($\medcirc$); Right: Ca$^{2+}$--Ca$^{2+}$ ($\medbullet$) and Ca$^{2+}$--Cl$^-$
($\medcirc$). $c_0 = 1.0$~mol\,dm$^{-3}$ , $c_\mathrm{HCl}^\mathrm{in} = 0.0436$~mol\,dm$^{-3}$,
$c_\mathrm{CaCl_2}^\mathrm{in} = 0.0522$~mol\,dm$^{-3}$. Symbols represent GCMC data while lines show
ROZ/HNC theory. $\lambda_\mathrm{B,0} = \lambda_\mathrm{B,1} = 7.14$~\AA.}
\label{fig:gr-ca}
\end{figure}

\begin{figure}[!h]
\centerline{\includegraphics[width=5.5cm]{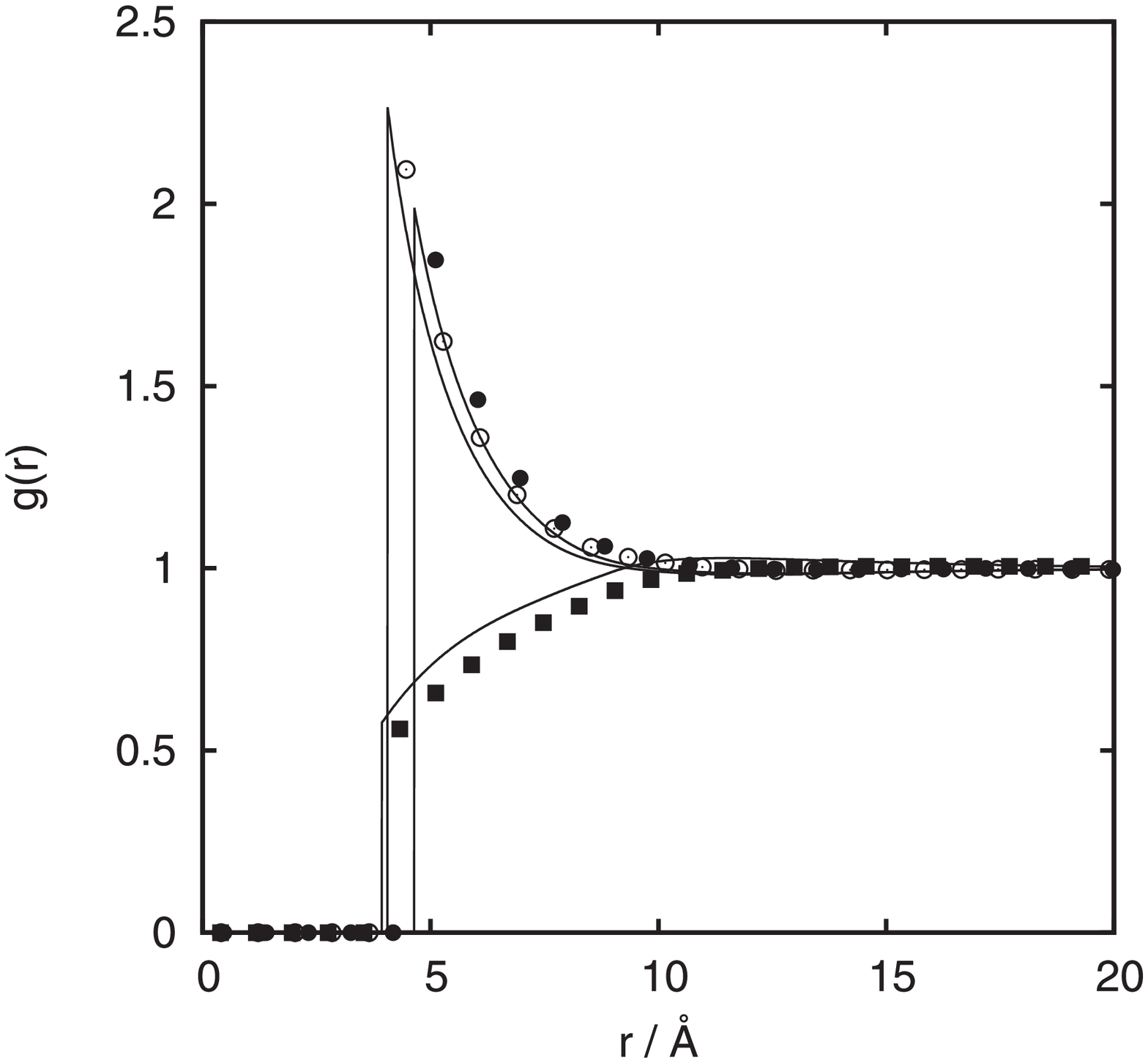}\hspace{1cm}
\includegraphics[width=5.5cm]{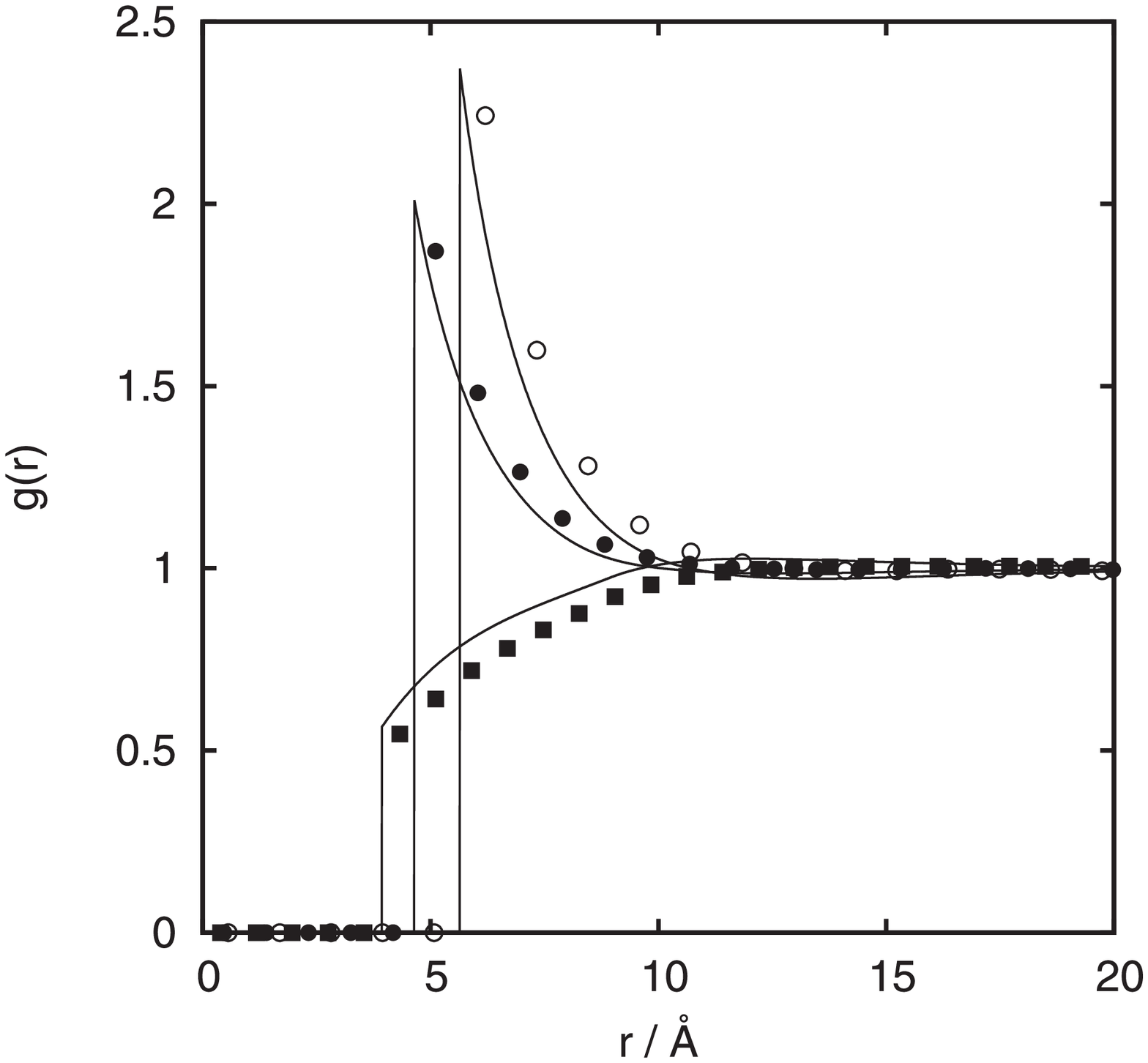}}
\caption{Fluid-matrix pair distribution functions (mixtures of HCl and NaCl (left) or HCl and CaCl$_2$ (right)). Left:
H$^+$--$\mathcal{M}$ ($\medbullet$), Na$^+$--$\mathcal{M}$ ($\medcirc$), and Cl$^-$--$\mathcal{M}$
($\blacksquare$), $c_\mathrm{HCl}^\mathrm{in} = 0.0325$~mol\,dm$^{-3}$, $c_\mathrm{NaCl}^\mathrm{in} =
0.2018$~mol\,dm$^{-3}$; Right: H$^+$--$\mathcal{M}$ ($\medbullet$), Ca$^{2+}$--$\mathcal{M}$ ($\medcirc$), and
Cl$^-$--$\mathcal{M}$ ($\blacksquare$), $c_\mathrm{HCl}^\mathrm{in} = 0.0436$~mol\,dm$^{-3}$,
$c_\mathrm{CaCl_2}^\mathrm{in} = 0.0522$~mol\,dm$^{-3}$. $c_0 = 1.0$~mol\,dm$^{-3}$. Symbols represent
GCMC data while lines show ROZ/HNC theory. $\lambda_\mathrm{B,0} = \lambda_\mathrm{B,1} = 7.14$~\AA.}
\label{fig:gr-matrix}
\end{figure}

To test the accuracy of the ROZ/HNC theory for the model of interest, we
compared the pair distribution functions and the mean activity
coefficients with the results of computer simulations. Figures~\ref{fig:gr-na} and \ref{fig:gr-ca} show the pair distribution functions
($++$ and $+-$) of the annealed ions, while figure~\ref{fig:gr-matrix}
shows the annealed ion-matrix distribution functions ($+\mathcal{M}$
and $-\mathcal{M}$). In all cases the concentration of the matrix was
$c_0 = 1.0$~mol\,dm$^{-3}$. Figure~\ref{fig:gr-na} and the left panel
of figure~\ref{fig:gr-matrix} apply to the mixture of HCl and NaCl, while
figure~\ref{fig:gr-ca} and the right panel of figure~\ref{fig:gr-matrix}
describe the distributions in the mixture of HCl and CaCl$_2$.
Symbols denote the GCMC and the lines denote the ROZ results.  Excellent
agreement of the two methods is obtained  for distribution
functions of hydrogen, while the agreement is slightly worse, but still
good, for the functions belonging to sodium  and calcium ions. Note that
the hydrogen ions are the ones compensating for the adsorbent
charge and are, therefore, present in an excess to other ions. A similarly
good agreement is obtained for the annealed ion-adsorbent pair
distribution functions shown in figure~\ref{fig:gr-matrix}.
\begin{table}[!h]
\caption{The same as in table~\ref{tbl:hclnacl}, but for HCl/CaCl$_2$ mixtures.}
\label{tbl:hclcacl2}
\footnotesize{
\begin{center}
\begin{tabular}{cccccccccc}
\hline\noalign{\smallskip}
\multicolumn{5}{c}{} & \multicolumn{2}{c}{$\gamma_\pm^\mathrm{HCl}$} & &
\multicolumn{2}{c}{$\gamma_\pm^\mathrm{CaCl_2}$}\\\noalign{\smallskip}
\hline\noalign{\smallskip}
$c_0$ & & $c_\mathrm{HCl}$ & $c_\mathrm{CaCl_2}$ & & GCMC & ROZ & & GCMC & ROZ\\\noalign{\smallskip}
\hline\noalign{\smallskip}
\multirow{7}{*}{1.0}
& & 0.00196	& 0.101   & & 1.060  &  1.085 & &  0.776 & 0.808\\
& & 0.00420	& 0.0952  & & 1.056  &  1.081 & &  0.773 & 0.806\\
& & 0.00804	& 0.0880  & & 1.051  &  1.078 & &  0.771 & 0.803\\
& & 0.0145	& 0.0791  & & 1.047  &  1.074 & &  0.767 & 0.801\\
& & 0.0253      & 0.0675  & & 1.043  &  1.070 & &  0.765 & 0.799\\
& & 0.0436      & 0.0522  & & 1.037  &  1.068 & &  0.761 & 0.798\\
& & 0.0751      & 0.0308  & & 1.034  &  1.067 & &  0.761 & 0.799\\\noalign{\smallskip}
\hline\noalign{\smallskip}
\multirow{8}{*}{0.5}
& & 0.00221       & 0.133 & &  0.863 & 0.870  & & 0.621  & 0.624\\
& & 0.00585       & 0.128 & &  0.863 & 0.869  & & 0.614  & 0.624\\
& & 0.0124        & 0.120 & &  0.861 & 0.868  & & 0.614  & 0.624\\
& & 0.0231        & 0.110 & &  0.861 & 0.867  & & 0.614  & 0.624\\
& & 0.0403        & 0.099 & &  0.860 & 0.867  & & 0.614  & 0.625\\
& & 0.0670        & 0.084 & &  0.861 & 0.869  & & 0.617  & 0.627\\
& & 0.108         & 0.064 & &  0.865 & 0.872  & & 0.620  & 0.631\\
& & 0.170         & 0.037 & &  0.870 & 0.879  & & 0.628  & 0.638\\\noalign{\smallskip}
\hline\noalign{\smallskip}
\multirow{9}{*}{0.1}
& &  0.00296        & 0.158 & & 0.763  &  0.763 & & 0.542  & 0.544\\
& &  0.0126         & 0.152 & & 0.764  &  0.763 & & 0.544  & 0.545\\
& &  0.0291         & 0.144 & & 0.765  &  0.765 & & 0.545  & 0.546\\
& &  0.0524	    & 0.135 & & 0.767  &  0.767 & & 0.547  & 0.548\\
& &  0.0827	    & 0.123 & & 0.770  &  0.770 & & 0.550  & 0.551\\
& &  0.121          & 0.120 & & 0.774  &  0.775 & & 0.554  & 0.555\\
& &  0.171          & 0.093 & & 0.780  &  0.780 & & 0.558  & 0.559\\
& &  0.236          & 0.071 & & 0.787  &  0.787 & & 0.564  & 0.565\\
& &  0.323          & 0.0411 & & 0.796  &  0.797 & & 0.572  & 0.572\\\noalign{\smallskip}
\hline\noalign{\smallskip}
\end{tabular}
\end{center}}
\vspace{-5mm}
\end{table}

The results for the mean activity coefficients, $\gamma_\pm$, in
aqueous mixtures of HCl with NaCl, and HCl with CaCl$_2$ in the matrix are
collected in tables~\ref{tbl:hclnacl} and \ref{tbl:hclcacl2}, respectively.
Note that in all cases, the particular electrolyte concentrations given,
are in equilibrium with the bulk solution of ionic strength $I^\mathrm{out} = 0.5
\sum_{i} z_i^2 c_i^\mathrm{out} = 0.5$~mol\,dm$^{-3}$. The agreement between the
ROZ/HNC and GCMC results is excellent for small matrix
concentration, but it deteriorates for larger $c_0$ values. At $c_0 =
1.0$~mol\,dm$^{-3}$, the mean activity coefficients calculated within
the ROZ/HNC theory are a bit too large, as already noticed in
previous papers
\cite{hribar2000,hribar2002,Luksic2007,Luksic2007a}.
Nevertheless, the differences discussed here are within the
estimated numerical errors of the two methods.

\subsection{The effect of the adsorbent capacity on the mean
  activity coefficient of the invading electrolyte.}

\begin{figure}[ht]
\centerline{\includegraphics[width=5.5cm]{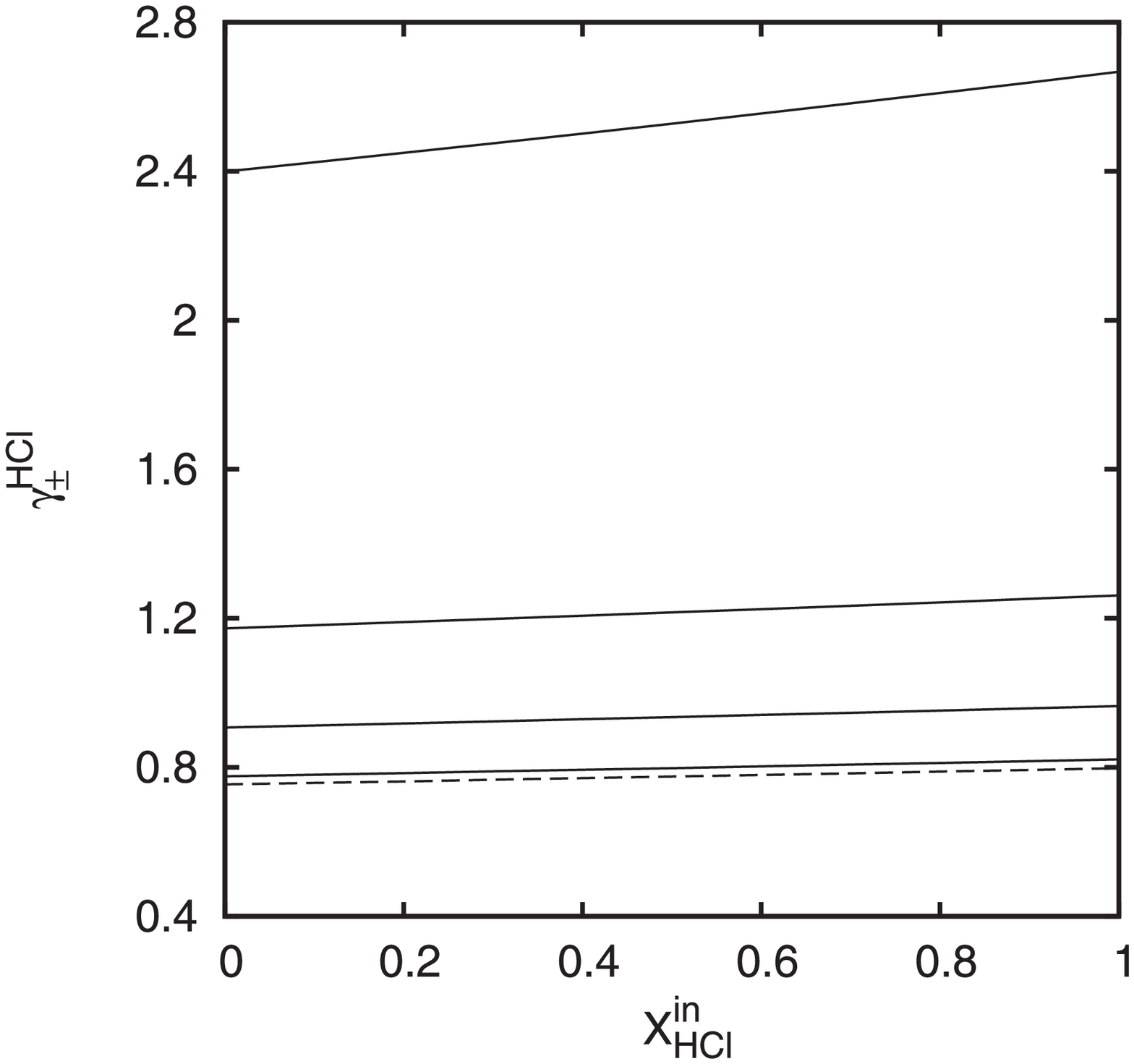}\hspace{1cm}
\includegraphics[width=5.5cm]{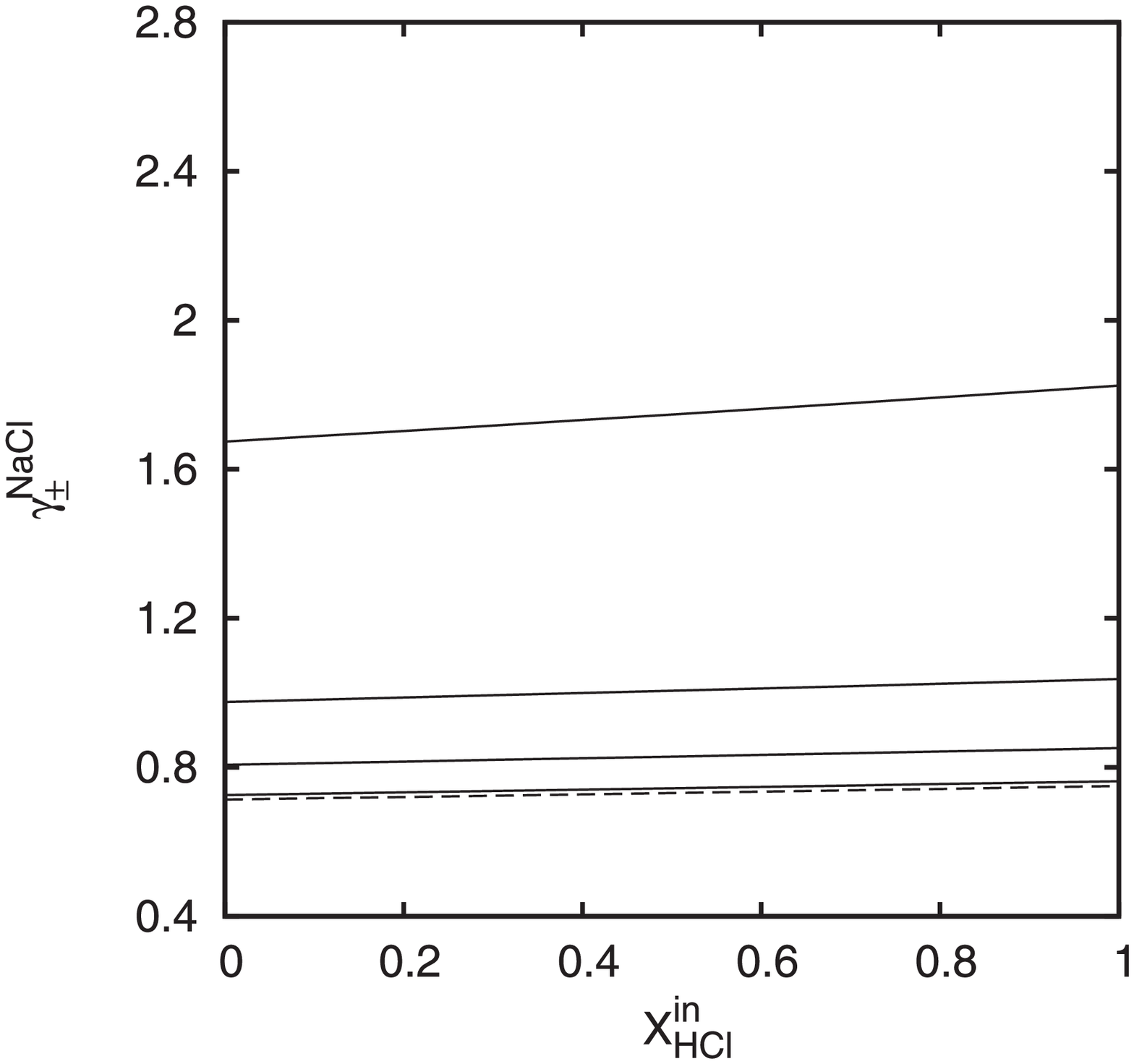}}
\caption{The mean activity coefficient as a function of the $X_\mathrm{HCl}^\mathrm{in}$ (mixtures of HCl and
NaCl), obtained using the ROZ/HNC theory. Left: HCl; Right: NaCl. From top to bottom (continuous lines): $c_0 =
2.0$, $1.0$, 0.5, and 0.1~mol\,dm$^{-3}$. Dashed line (obtained using OZ/HNC) represents the bulk value.
$I^\mathrm{in} = 0.5$~mol\,dm$^{-3}$,  $\lambda_\mathrm{B,0} = \lambda_\mathrm{B,1} = 7.14$~\AA.}
\label{fig:gamma-na}
\end{figure}

\begin{figure}[ht]
\centerline{\includegraphics[width=5.5cm]{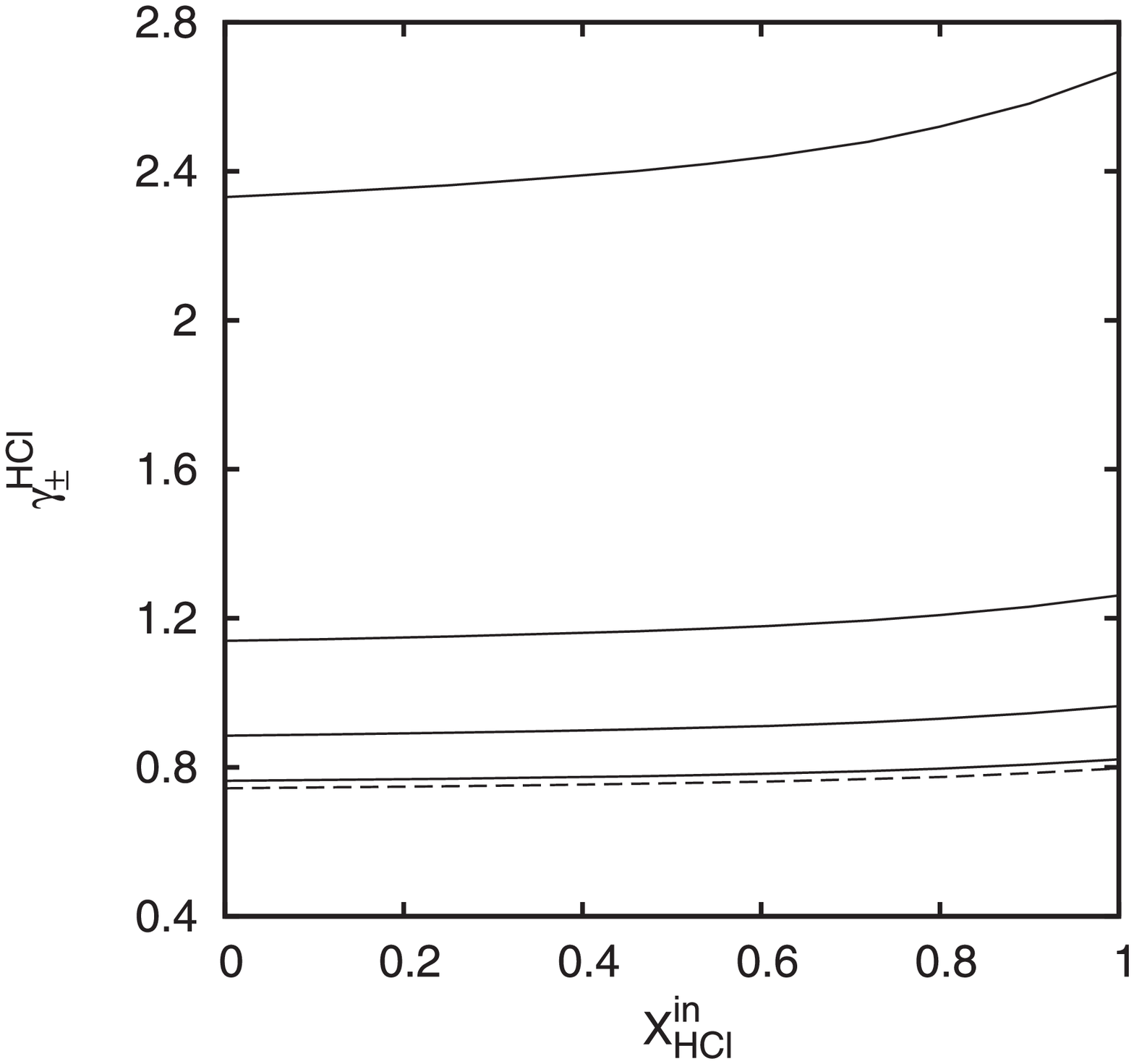}\hspace{1cm}
\includegraphics[width=5.5cm]{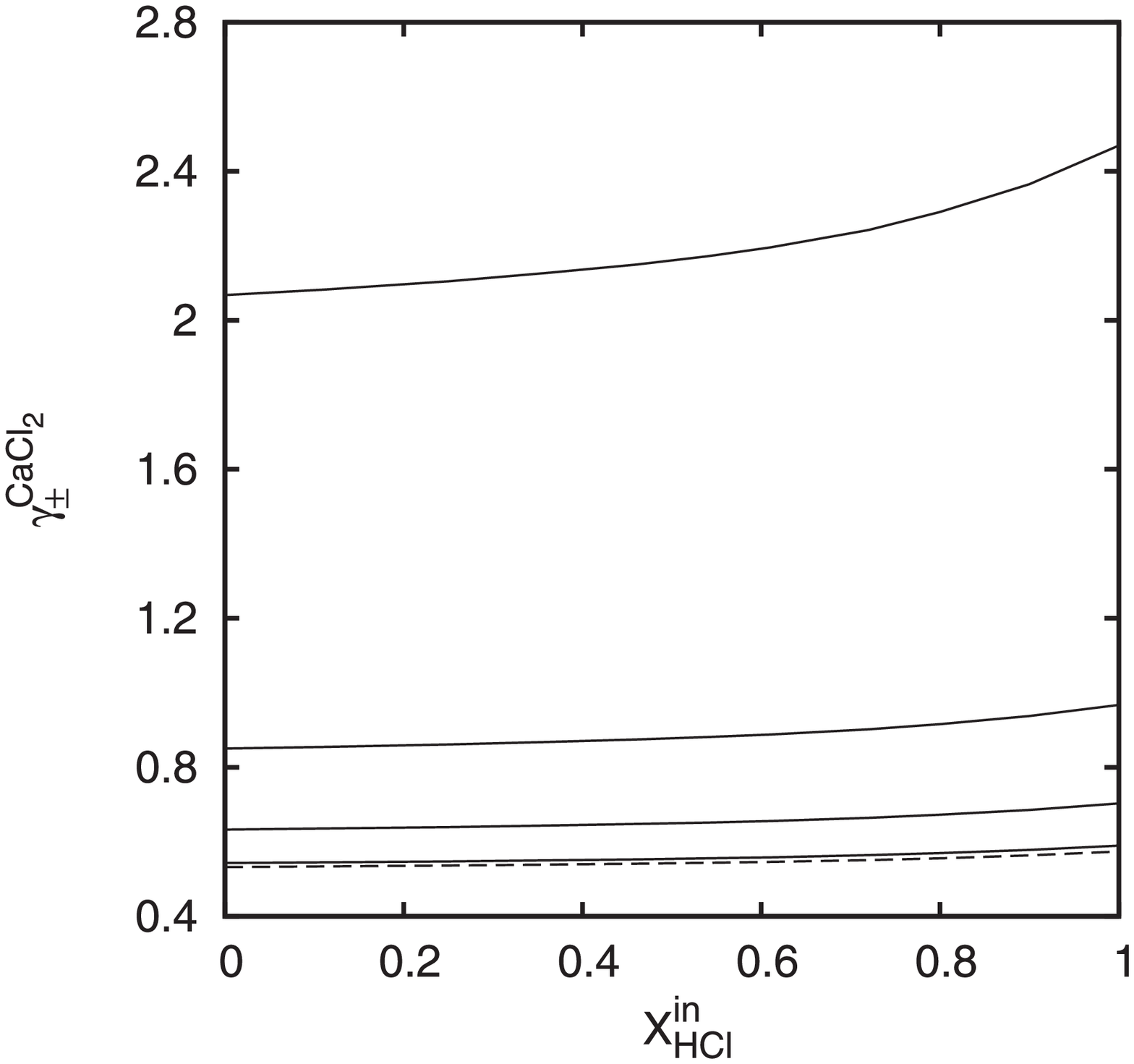}}
\caption{The same as in figure~\ref{fig:gamma-na}, but for HCl/CaCl$_2$ mixtures.}
\label{fig:gamma-ca}
\end{figure}

The ROZ/HNC method was used to investigate the effect of the
adsorbent capacity, given here as the matrix concentration,
on the mean activity coefficient of the invading electrolyte. Note that
while the ionic strength of the annealed electrolyte mixture within the
matrix (denoted by superscript ``in'') was kept constant, $I^\mathrm{in} = 0.5$
mol\,dm$^{-3}$, for all cases examined here, the composition given by
\begin{equation}
X_\mathrm{HCl}^\mathrm{in} = \frac{c_\mathrm{HCl}^\mathrm{in}}{c_\mathrm{HCl}^\mathrm{in} +
c_\mathrm{Z}^\mathrm{in}} = 1 - X_\mathrm{Z}^\mathrm{in}
\label{eq:x}
\end{equation}
varied. Z in equation (\ref{eq:x}) denotes either NaCl or CaCl$_2$. Concentrations of the matrix obstacles considered were: $c_0 = 0.1$, 0.5, 1.0, and
2.0~mol\,dm$^{-3}$. The ROZ/HNC results are shown in figures~\ref{fig:gamma-na} (HCl/NaCl), and \ref{fig:gamma-ca}
(HCl/CaCl$_2$) by continuous lines, the dashed lines (OZ/HNC) show
the mean activity coefficients in the bulk solution with $I = 0.5$~mol\,dm$^{-3}$.
\begin{figure}[!ht]
\centerline{\includegraphics[width=5.5cm]{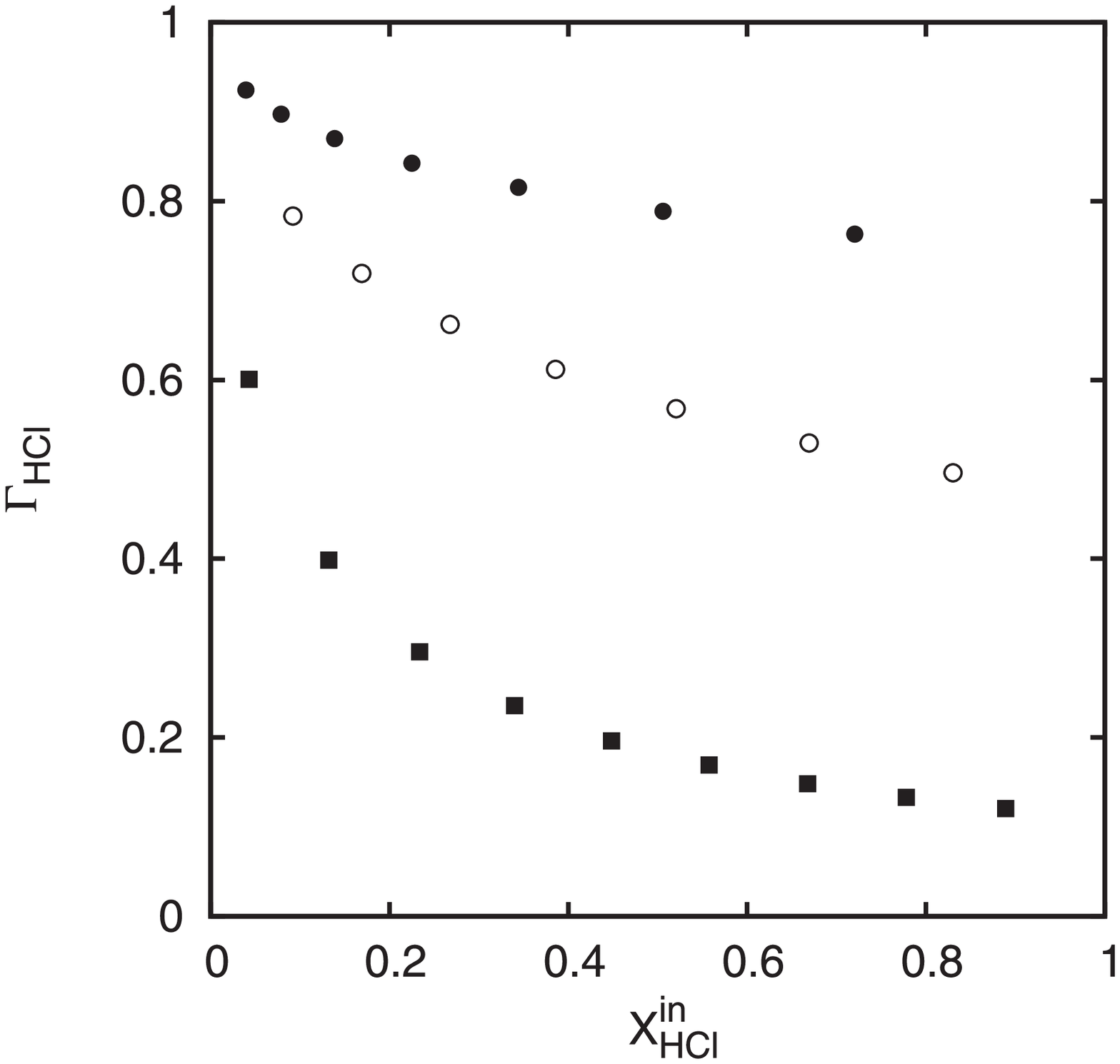}\hspace{1cm}
\includegraphics[width=5.5cm]{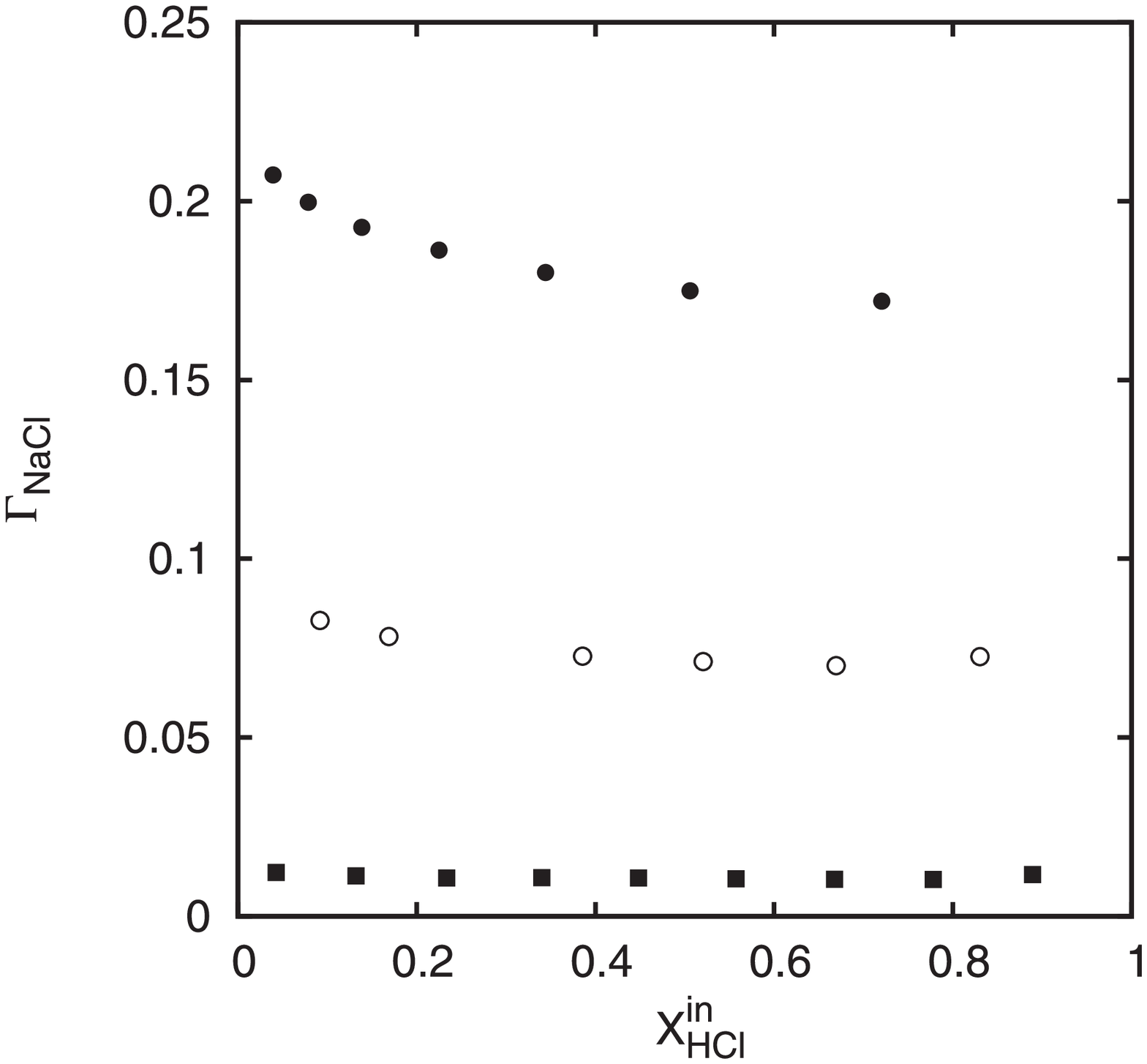}}
\caption{Donnan exclusion coefficient as a function of $X_\mathrm{HCl}^\mathrm{in}$ (mixtures of HCl and NaCl), obtained with the GCMC simulations. Left: HCl; Right:
NaCl. $c_0 = 1.0$ ($\medbullet$), 0.5 ($\medcirc$), and 0.1 ($\blacksquare$)~mol\,dm$^{-3}$. $I^\mathrm{out}=0.5$~mol\,dm$^{-3}$,
$\lambda_\mathrm{B,0} = \lambda_\mathrm{B,1} = 7.14$~\AA.}
\label{fig:donnan-na}
\end{figure}

As the matrix concentration (capacity of the adsorbent) increases, the
mean activity coefficients of all the species in the solutions increase
from values smaller than unity to values higher than unity, as
previously observed for a single electrolyte adsorbed in charged
matrix \cite{Luksic2007,Luksic2007a}. The matrix charge repels the
co-ions of the invading electrolyte mixture, which causes the exclusion of
the electrolyte on the  whole from the matrix. Concentration of a
particular electrolyte is smaller in the matrix than in the corresponding
equilibrium bulk solution. This can be clearly seen from the values of
the Donnan exclusion coefficient, $\Gamma_i$, for a particular salt
$i$ (HCl, NaCl, or CaCl$_2$). $\Gamma_i$ is defined as
\begin{equation}
\Gamma_i = \frac{c_i^\mathrm{out}  -  c_i^\mathrm{in}}{c_i^\mathrm{out}} \,.
\end{equation}

\begin{figure}[ht]
\centerline{\includegraphics[width=5.5cm]{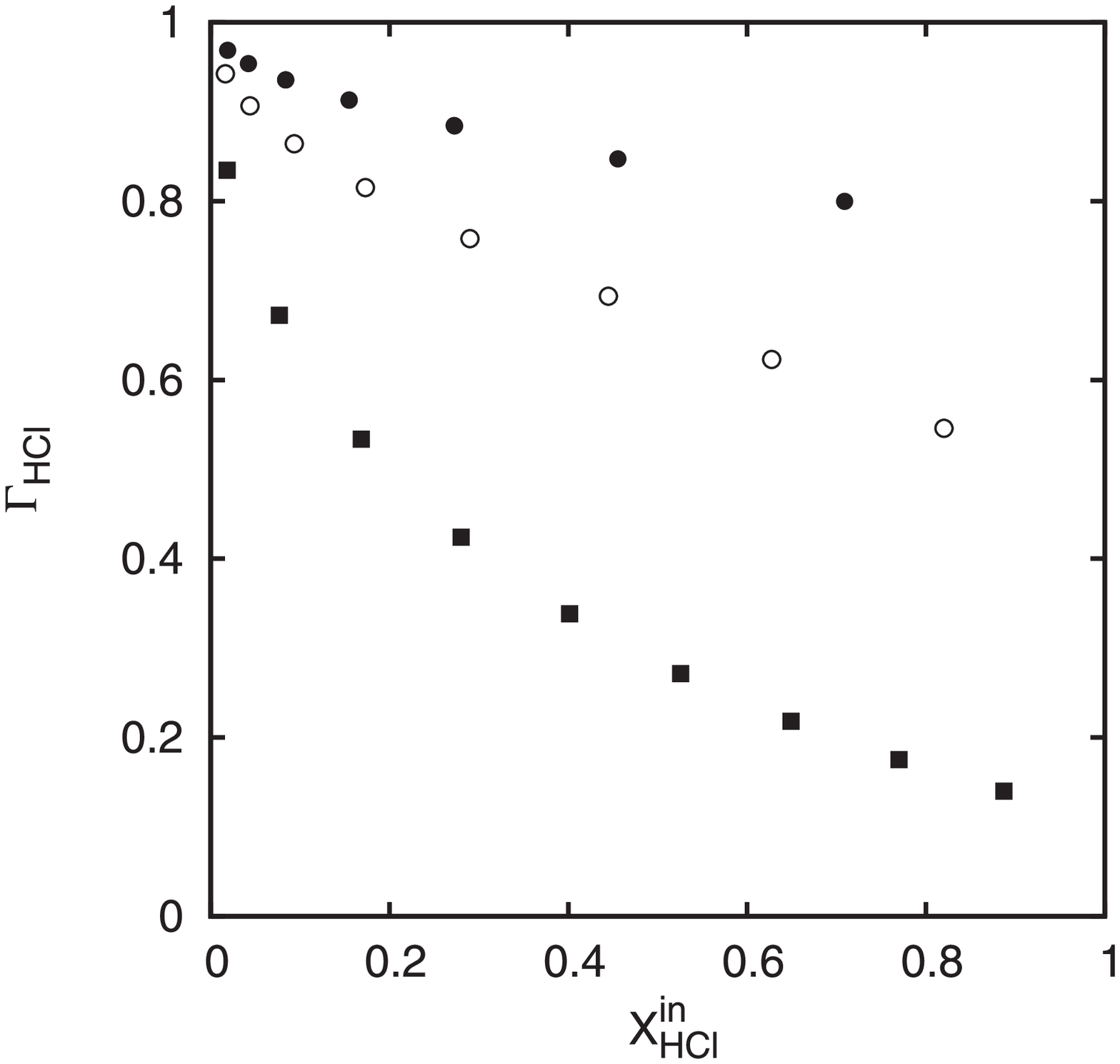}\hspace{1cm}
\includegraphics[width=5.5cm]{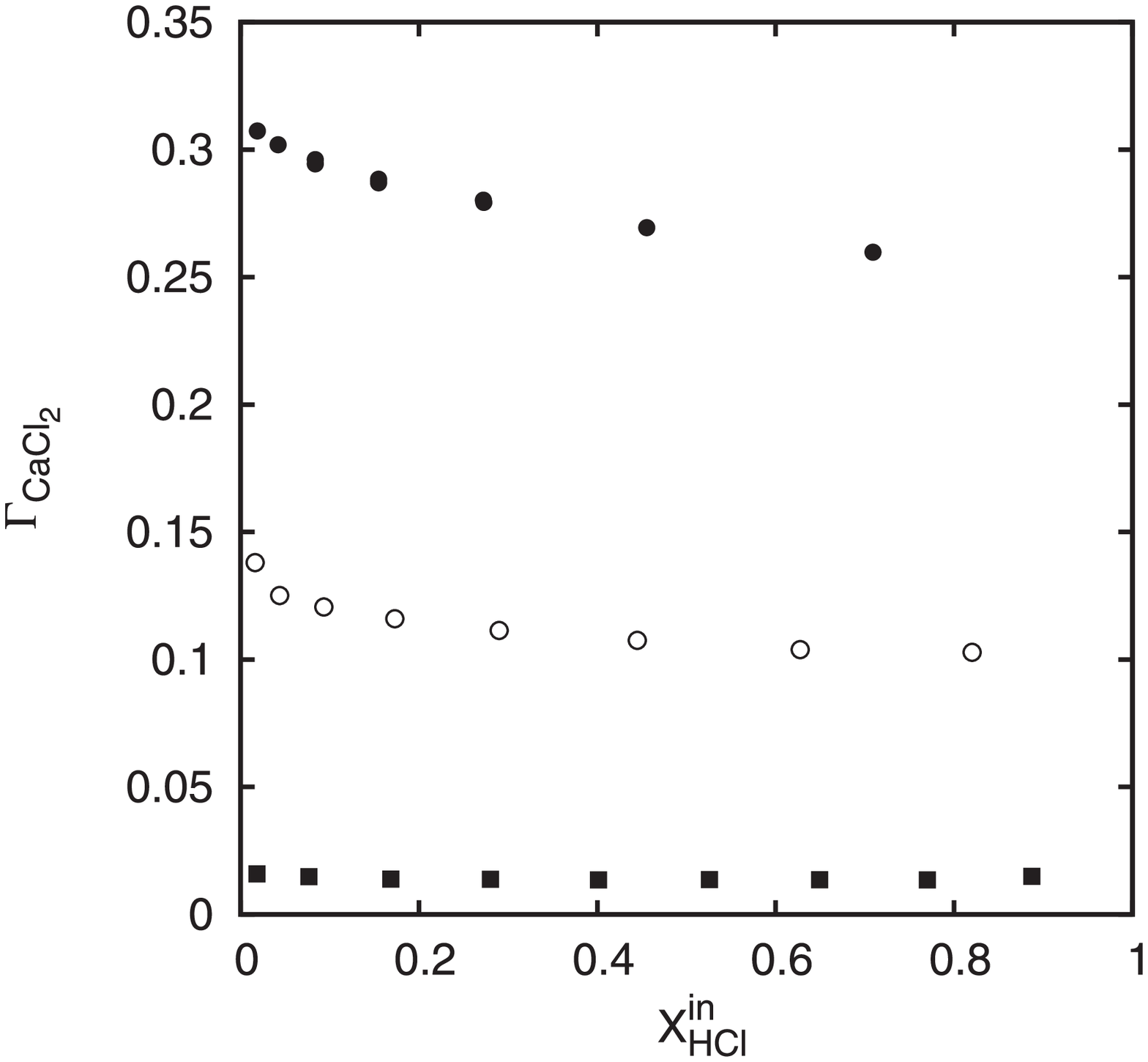}}
\caption{The same as in figure~\ref{fig:donnan-na}, but for mixtures of HCl and CaCl$_2$.}
\label{fig:donnan-ca}
\end{figure}

The results for the Donnan exclusion coefficients, calculated by the
GCMC method (ionic strength of the bulk electrolyte was kept
constant, $I^\mathrm{out}=0.5$~mol\,dm$^{-3}$), are shown in
figures~\ref{fig:donnan-na} (HCl/NaCl) and \ref{fig:donnan-ca}
(HCl/CaCl$_2$). As indicated by the behaviour of the mean activity
coefficients (figures~\ref{fig:gamma-na} and \ref{fig:gamma-ca}), the
annealed electrolytes are excluded from the adsorbent (positive
$\Gamma$ values) in all cases. The exclusion is stronger for higher
matrix concentration (i.e. higher adsorbent capacity, or higher
charge), and depends on the composition of the invading electrolyte.
The exclusion coefficient for all electrolytes in the mixture decreases with
the increasing fraction of the HCl in the solution. While the
$\Gamma_\mathrm{HCl}$ approaches the value one as $X_\mathrm{HCl}^\mathrm{in} \rightarrow 0$ (i.e. for no exclusion), the exclusion coefficient of the
other electrolyte in the mixture approaches a somewhat lower value.
Such a result has previously been obtained for
electroneutral adsorbents \cite{Luksic2009a}. The
difference in the behaviour of the two electrolyte components is caused by the modelling of the system (see section~2),
mimicking the H$^+$-ion-exchange resin. The matrix charge is namely
``neutralized'' with hydrogen ions, which are accordingly always
present in the annealed electrolyte mixture.

\subsection{The ion-exchange isotherms}

Due to a different degree of adsorption (exclusion), the
concentration ratio of the competing ions in the adsorbent (cations) is
generally different from that in the bulk solution. In other words, the
adsorbent ``selects'' one species in preference to the other. The study
of selectivity was one of the main goals of this work.
\begin{figure}[!ht]
\centerline{\includegraphics[width=5.5cm]{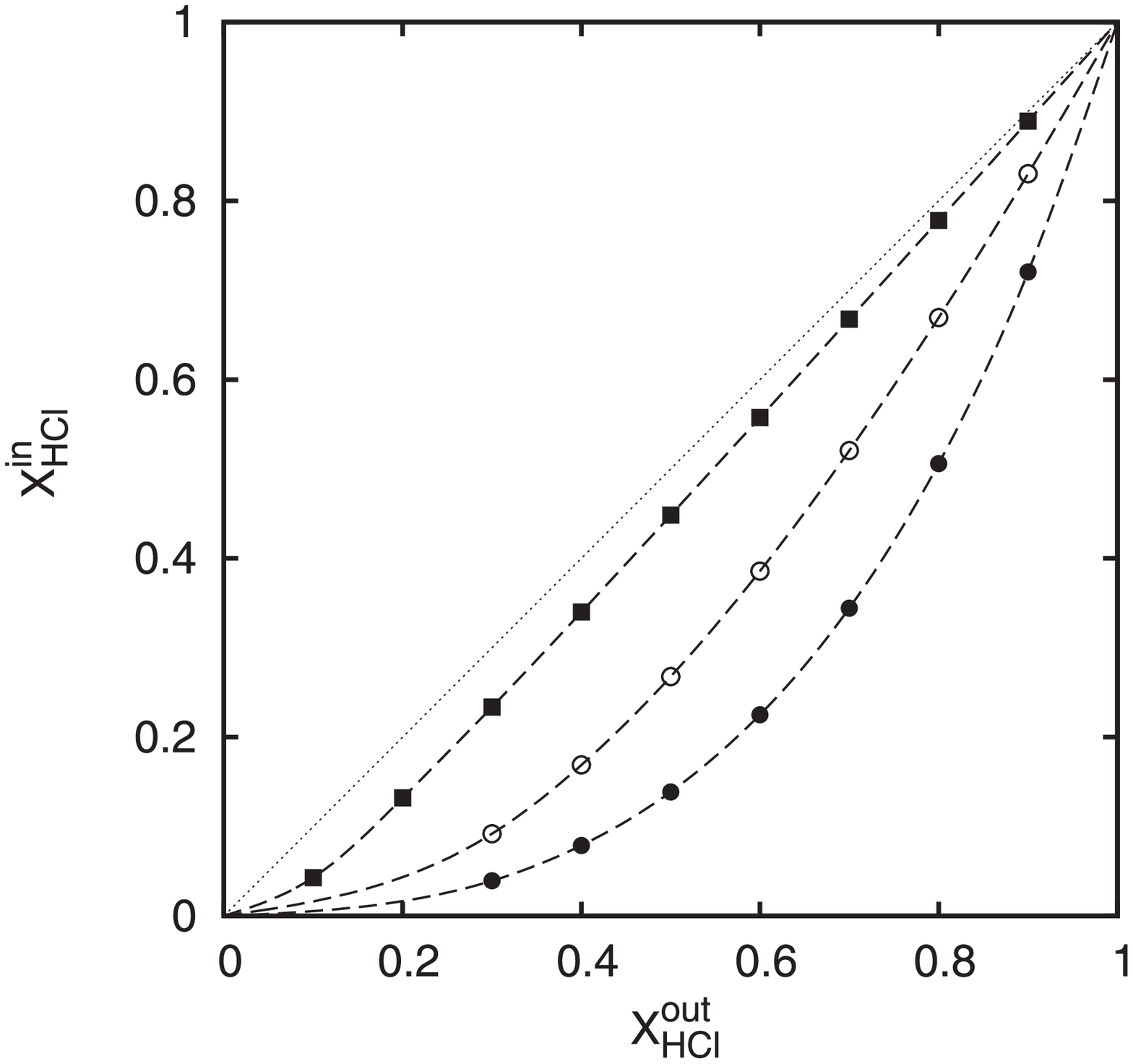}\hspace{1cm}
\includegraphics[width=5.5cm]{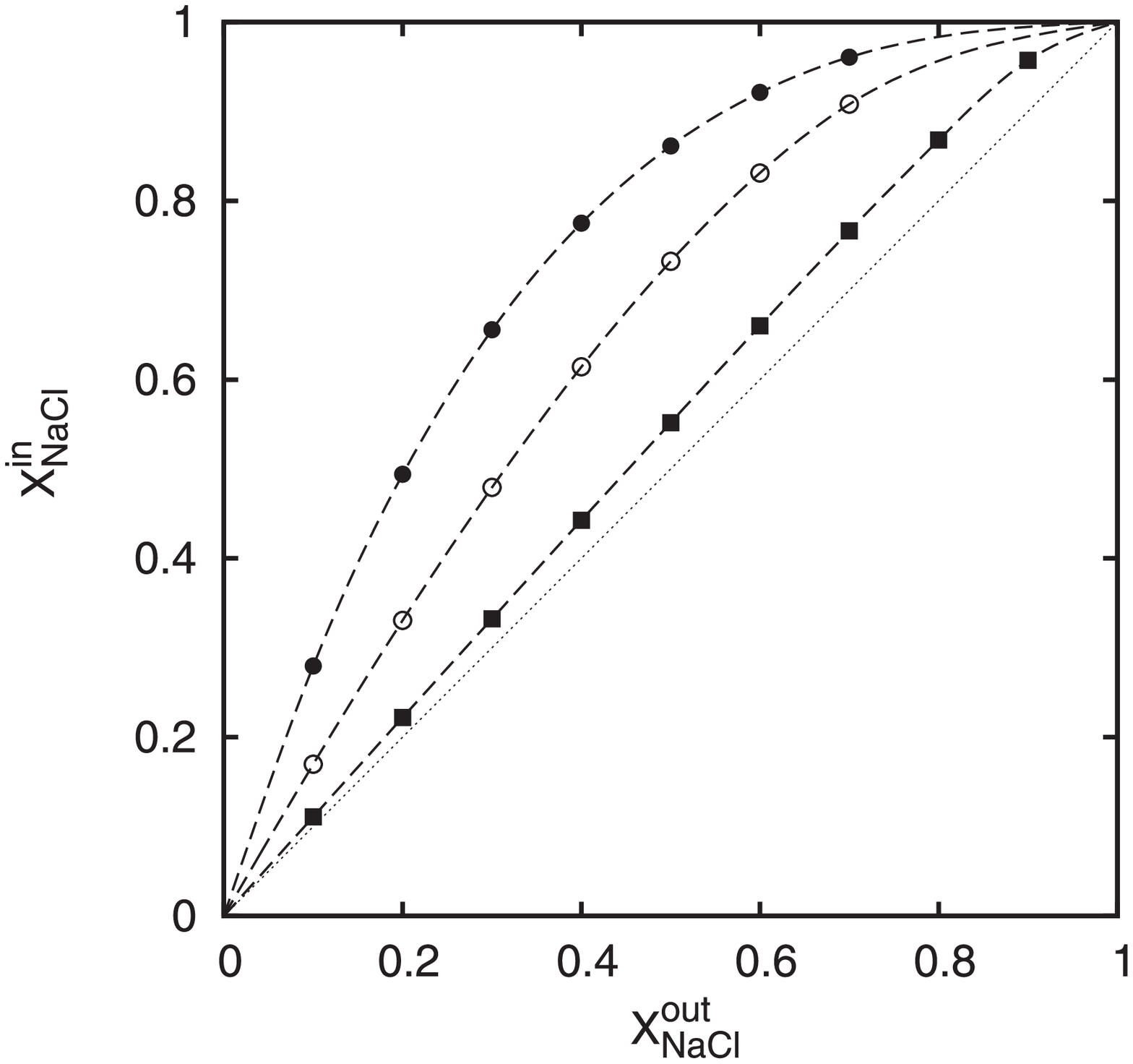}}
\caption{Ion-exchange isotherms (mixtures of HCl and NaCl), obtained with the GCMC simulations. Left: HCl; Right:
NaCl. $c_0 = 1.0$ ($\medbullet$), 0.5 ($\medcirc$), and 0.1 ($\blacksquare$)~mol\,dm$^{-3}$.
$\lambda_\mathrm{B,0} = \lambda_\mathrm{B,1} = 7.14$~\AA. Lines are guides for the eye.}
\label{fig:xinout-na}
\end{figure}
Here we
present the results in the form of ion-exchange isotherms, showing the
mole fraction of one electrolyte component within the adsorbent,
$X^\mathrm{in}$, as a function of the composition of the equilibrium
external solution, $X^\mathrm{out}$. The ion-exchange isotherms are
presented in figure~\ref{fig:xinout-na} (HCl/NaCl mixture) and figure~\ref{fig:xinout-ca} (HCl/CaCl$_2$ mixture). Symbols show the grand
canonical MC results at different matrix concentrations, and the
connecting lines merely serve to guide the eye. The diagonal lines
apply to a hypothetical case with no preference to the adsorbent.
\begin{figure}[!ht]
\centerline{\includegraphics[width=5.5cm]{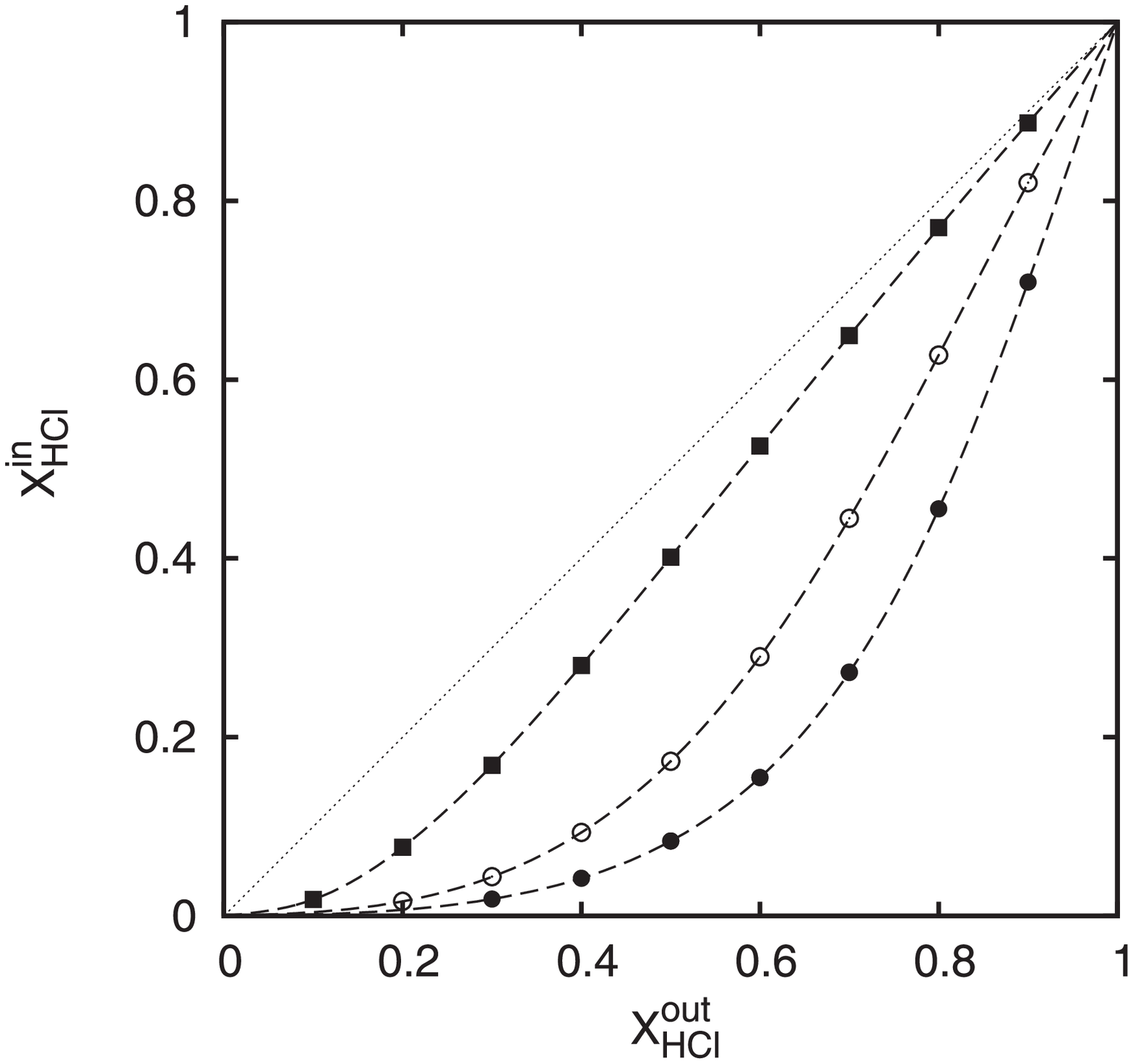}\hspace{1cm}
\includegraphics[width=5.5cm]{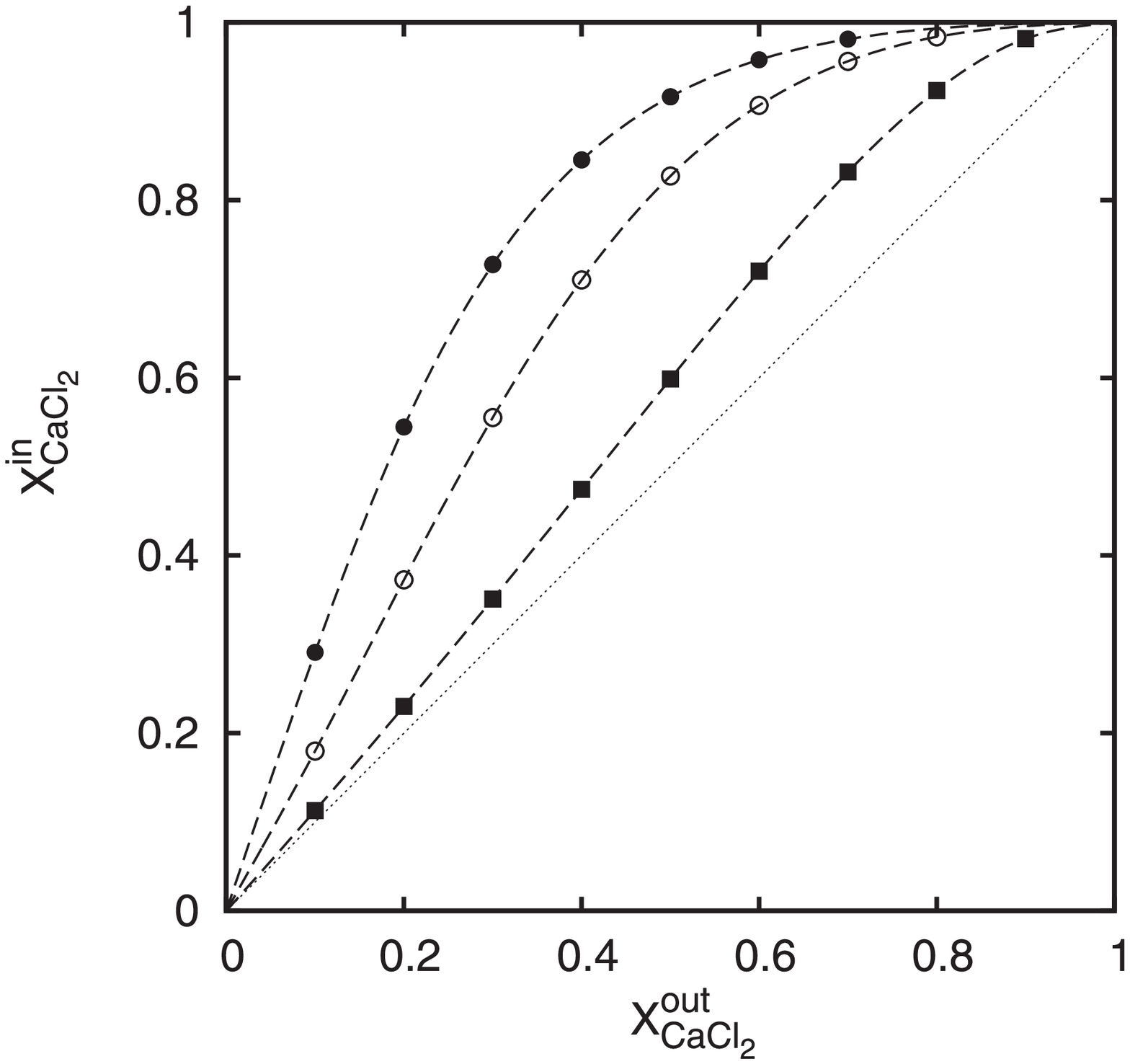}}
\caption{The same as in figure~\ref{fig:xinout-na}, but for mixtures of HCl and CaCl$_2$.}
\label{fig:xinout-ca}
\end{figure}

As noticed before \cite{Luksic2009a,Trefalt2007}, the selectivity
increases with the increased capacity of the adsorbent (increased matrix
concentration). In all cases, the mole fraction of the HCl solution in
the adsorbent is smaller than in the external solution (left panels of
figures~\ref{fig:xinout-na} and \ref{fig:xinout-ca}), the opposite is true
for the NaCl or CaCl$_2$ (right panels of figures~\ref{fig:xinout-na}
and \ref{fig:xinout-ca}), respectively. The selectivity strongly depends
on the composition of the solution as well. For the component that is
preferentially adsorbed in the matrix (NaCl and CaCl$_2$ in our
case), the selectivity increases with an increasing fraction of the
component in the solution, while the opposite is true for the other
component. Similar trends were observed experimentally
\cite{Helferich,Vlachy1990}. Due to a stronger electrostatic interaction
between the matrix charges and the ion with higher charge density
(Na$^+$ or Ca$^{2+}$, respectively), the latter ions are excluded
from the matrix to a smaller extent.

\section{Conclusions}

The model of ion-exchange resin was studied using the ROZ/HNC theory and
Monte Carlo simulation in
grand canonical ensemble. Theoretical results for the pair
distribution functions were found to be in good agreement with
computer simulation results; it is confirmed that ROZ theory represents
a viable alternative to computer simulations. For all the electrolyte
mixtures and matrix concentrations
studied here, the electrolyte components are excluded from the
adsorbent phase. This is most often explained as a consequence of electrostatic repulsion
between matrix charges and co-ions from the electrolyte component.
However, the ionic species are adsorbed to a
different amount. Consistently with experimental observations,
the ions with higher charge density (sodium and calcium ions in our
case) adsorb to a greater extent.

\section*{Acknowledgements}

The authors appreciate the financial support of the Slovenian
Research Agency via Program P1--0201 and the Project J1--4148.


\ukrainianpart

\title{Моделювання іонно-обмінної взаємодії в нанопористих матеріалах}
  \author{М.~Лукшіч, В.~Влахи, Б.~Xрібар-Лі}
  \address{Унiверситет Любляни, факультет хiмiї та хiмiчних технологiй, Ашкерчева 5, SI--1000, Любляна, Словенiя}

\makeukrtitle

\begin{abstract}
\tolerance=3000%
Використовуючи теорію реплікованого рівняння Орнштейна-Церніке і метод
Монте Карло у великому канонічному ансамблі, досліджується розподіл
двокомпонентної електролітичної суміші між модельним адсорбентом і
водним розчином електроліту.  Компоненти електроліту моделюють суміші
HCl/NaCl і HCl/CaCl$_2$.  Матриця, заповнена примітивною моделлю
електролітичної суміші, була сформована з моновалентних негативно
заряджених сферичних частинок-перешкод.  Розчин розглядався як
неперервний діелектрик з властивостями чистої води. Порівняння парних
функцій розподілу (отримане обома методами) між різними іонними
сортами вказують на добре узгодження між результатами теорії репліки
Орнштейна-Церніке і машинних розрахунків.  Серед термодинамічних
властивостей, розраховано коефіцієнт середньої активності компонентів
електроліту в матриці. Запропоновано просту модель для іонно-обмінної
взаємодії. Розрахунок селективності дав якісне узгодження з такими
експериментальними даними: (i) селективність зростає з ростом ємності
адсорбента (концентрації матриці), (ii) адсорбент проявляє більшу
селективність до іону з вищою зарядовою густиною, якщо його частка в
суміші є меншою.

\keywords адсорбція, змішані електроліти, селективність, Монте
Карло, теорія репліки Орнштейна-Церніке

\end{abstract}

\end{document}